\documentclass{aa}
\usepackage[colorlinks=true,allcolors=black]{hyperref}
\usepackage{pdflscape}
\usepackage{multirow}
\usepackage{longtable}
\usepackage{graphicx}
\usepackage{rotating}
\usepackage{booktabs}
\usepackage{txfonts}
\newcommand{\dsct}{$\delta$~Scuti }

\newcommand{\HAS}{\hyperlink{cite.Hasanzadeh2021}{H21}}

\usepackage{array}
\usepackage{amssymb}
\usepackage{url}
\usepackage{multirow}
\usepackage{makecell}
\usepackage{adjustbox}

\begin{document}

   \title{Statistical analysis of asteroseismic indices and stellar parameters of TESS-observed \dsct stars}
   \author{C.\,Lv
          \inst{1,2}
          \and
         A.\,Esamdin
          \inst{1,2}
          \and
         A.\,Hasanzadeh
          \inst{3,4}
          \and
         M.\,Ghazinejad
          \inst{5}
          \and
         J.Pascual-Granado
          \inst{6}
          \and
         G.\,M. Mirouh
          \inst{7}
          \and
         R.\,Karimov
          \inst{8}
          }

   \institute{Xinjiang Astronomical Observatory, Chinese Academy of Sciences, Urumqi, Xinjiang 830011, People's Republic of China\\
              \email{aliyi@xao.ac.cn}
        \and
             School of Astronomy and Space Science, University of Chinese Academy of Sciences, Beijing 100049, People's Republic of China\\
        \and
             Centre for Fusion, Space and Astrophysics, Department of Physics, University of Warwick, Coventry CV4 7AL, UK\\
        \and
             Institute of Geophysics, University of Tehran, Tehran, Iran\\
        \and
             Department of Physics, Faculty of Science, University of Qom, Iran\\
        \and Instituto de Astrof\'isica de Andaluc\'ia - CSIC, 18008 Granada, Spain\\
        \and Departamento de Física Teórica y del Cosmos, Universidad de Granada, Campus de Fuentenueva s/n, 18071 Granada, Spain\\
        \and
             Ulugh Beg Astronomical Institute, Uzbekistan Academy of Sciences, Tashkent 100052, Uzbekistan\\
             }
  \abstract
{Over the last few years, \dsct stars have been at the center of the attention of the asteroseismology community thanks to the derivation of seismic indices connected to stellar parameters. The statistical analysis of the wealth of data offered by a large space survey such as the Transiting Exoplanet Survey Satellite (TESS), the identification of new \dsct stars, and the correlation between asteroseismic indices and stellar parameters in the resulting sample are therefore of utmost interest.}
   {The goal of our study is to analyze the statistical properties of stellar parameters and characterize the asteroseismic indices of \dsct stars observed in TESS cycle 4.}
   {We used TESS 2 min cadence photometric data and the corresponding Fourier transform to identify \dsct stars. The asteroseismic indices for these stars were determined using an empirical relation and a 2D autocorrelation method.}
   {We discovered 765 \dsct stars from the data obtained by the TESS mission, from Sectors 40 to 55, corresponding to cycle 4 and observed with a 2 min cadence. Of these stars, 179 \dsct stars have low-resolution spectral parameters from LAMOST. We first analyzed the relation between pulsation and stellar parameters from TESS observations and the distribution of \dsct stars with two different stellar parameters, TESS Input Catalog (TIC) and  Large Sky Area Multi-Object Fiber
Spectroscopic Telescop (LAMOST), within the classical instability strip. Most of the stars lie within the instability strip and follow the period-luminosity relation of \dsct stars. Additionally, the majority of the stars exhibit pulsation properties consistent with those expected for \dsct stars, including periods falling within the typical range, amplitudes at the millimagnitude level, and fundamental parameters such as spectral type, effective temperature, log g, and luminosity that match the characteristics of \dsct stars. This confirms the reliability of the \dsct stars we have identified. We subsequently obtained the large frequency separation ($\Delta\nu$), $\nu_{\mathrm{max}}$, and $\nu({\rm Amax})$ for 179 \dsct stars with LAMOST parameters by using an empirical relation and a 2D autocorrelation method, and obtained the relations between these asteroseismic indices. These stars will provide significant support for a deeper study of the internal structure and evolution of stars.}
   {}
   \maketitle
\section{Introduction}
\label{sect:Introduction}

In recent years, we have seen a rapid development of photometric space telescopes such as the Microvariability and Oscillations of STars \citep[MOST,][]{2003PASP..115.1023W}; Convection, Rotation, and planetary Transits \citep[CoRoT,][]{2009A&A...506..411A}; $Kepler$ (e.g., \citealp{2010PASP..122..131G,2010ApJ...713L..79K}); and TESS \citep{Ricker2015}, which has allowed us to obtain a wealth of high-resolution data. Studies utilizing these databases can investigate stellar structure and evolution and identify numerous targets that are challenging to detect through ground-based observations.
The
\dsct stars are A-F stars pulsating with pressure modes, typically located in the overlap of the main sequence and pulsating instability strip in the Hertzsprung$-$Russell (HR) diagram. There is also a fraction of \dsct stars in the pre-main-sequence (Pre-MS) evolutionary stage. The first detection of \dsct oscillations in a Pre-MS star was reported by \citet{Kurtz1995}, and analyzing the stellar pulsations of such stars can offer constraints on the physics of protostellar accretion (e.g., \citealp{Zwintz2014,Steindl2021}). \dsct stars typically have masses in the $1.5-2.5\ $M$_{\odot}$ range; thus, they lie on the Kraft break (e.g., \citealp{Kraft1967,Royer2009,Avallone2022}). This represents the transition between slowly rotating low-mass stars with a radiative core and a thick convective envelope and rapidly rotating intermediate-mass stars with a convective core and a major radiative envelope (e.g., \citealp{Kraft1967,Royer2009,Avallone2022}).

Most \dsct stars are main-sequence or subgiant stars with no, or a thin, convective envelope. The differences in the structure of stars offer opportunities to investigate various areas of physics such as pulsation, rotation, magnetic fields, and chemical peculiarities (e.g., \citealp{Murphy2015,2015MNRAS.447.3264S,2021MNRAS.500.1992T}). Hence, the examination of \dsct star is important for testing theoretical models of stellar evolution. These stars manifest numerous pulsation frequencies associated with different modes of pulsation, primarily radial and non-radial modes (e.g., \citealp{Breger2000,Uytterhoeven2011}), which are generally excited by the $\kappa$ mechanism \citep{ASTERO}. The spectral types of \dsct stars usually span from A2 to F2, and their effective temperature falls within the range of $6300 \leq$ T$_{\rm eff} \leq 8600$~K \citep{Uytterhoeven2011}. These stars exhibit pulsation modes mainly characterized as low-radial-order ($n$), low-degree ($l$) pressure ($p$) modes (e.g., \citealp{1998A&A...335..549V,ASTERO,lv2021AJ,lv2022AJ,lv2022APJ}). Investigating binary systems containing \dsct stars is also valuable as it provides more accurate stellar parameters. Therefore, such systems are excellent targets for asteroseismic studies, which can help advance our understanding of stellar structure and evolution (e.g., \citealp{2019ApJ...885...46G,Murphy2020,lv2021RAA,2021MNRAS.505.3206M,chen2022}).

\citet{Bowman2016} conducted an extensive analysis on the modulation of both amplitude and phase within the frequency spectra of 983 \dsct stars. These stars were observed over a span of four years through the $Kepler$ mission. The results revealed that more than 60\% of these celestial objects displayed significant fluctuations in both amplitude and phase. The study systematically investigated the root causes of these variations in amplitude and phase, encompassing factors such as unresolved frequencies, whether they were mode frequencies or nonlinear combination frequencies, the presence of three-mode resonances, variations in driving and damping mechanisms, and the intricate exchange of energy among coupled modes.

Since the launch of the first asteroseismic space missions, the power spectra of \dsct stars have resisted mode identification for a substantial portion of their frequencies, which is a crucial step in the process of forward modeling. A single star can exhibit the excitation of hundreds of modes, including radial, nonradial, and mixed modes, as well as combination frequencies resulting from nonlinear mode interactions \citep{ASTERO}, but even the most sophisticated models are incapable of fitting the full number and range of frequencies observed in some cases (e.g., \citealp{Garc2009, Balona2023}). In particular, in \citet{Poretti2009} a flat power excess made of low-amplitude peaks is shown at low frequencies. Several plausible mechanisms to explain this include a cascade bug in the pre-whitening procedure \citep{Balona2014}, an inconsistency in the analysis due to the non-harmonic nature of the signal (e.g., \citealp{PG15, DF18}), mode coupling (e.g., \citealp{Buchler1997, BarceloForteza2015}), and a background signal due to granulation \citep{Kallinger2010}. Furthermore, most \dsct stars are characterized as moderate-to-rapid rotators, which causes the rotationally split multiplets of non-radial modes to have uneven separations (e.g., \citealp{reese22,Kurtz2022}) and the modes to assume complex geometries (e.g., \citealp{ballot13,Mirouh2022}). This discrepancy between the multiplet separation and the mode frequency separation complicates efforts to identify specific modes. The lack of a precise identification of pulsation modes within the frequency spectra poses a significant obstacle when attempting to construct accurate asteroseismology models for these stars. In such cases, asteroseismic indices may offer valuable insights and assistance in the process of mode identification (e.g., \citealp{2020Natur.581..147B}, \citealp{Hasanzadeh2021}; hereafter \HAS ).

The use of scaling relations based on asteroseismic indices and global stellar properties is a common approach to determining the fundamental parameters of stars (e.g., \citealp{Brown1991,Kallinger2010,Coelho2015,Rodrigues2017,Hekker2020}, \HAS ). However, validating the accuracy and reliability of these relations is a crucial step (e.g., \citealp{Huber2017}, \HAS ). The stellar density and large separation were determined from the characteristics of binaries containing a \dsct component by \citet{Garc2015,Garc2017}, who found that the relationship between these two parameters deviated slightly from the theoretical expectations obtained from the asymptotic theory of \citet{Tassoul80}.
\citet{Mirouh2019} found large separations in models to be about 0.9 times the predictions derived from the asymptotic theory, a phenomenon that was observed in 60 young \dsct stars by \citet{2020Natur.581..147B} with large separations of 0.85 times their asymptotic value. This is due to the radial orders of the detected modes being too low for the large separation to reach its asymptotic value \citep{Garc2009}.
Meanwhile, a new relationship between large separation and average density was obtained in (e.g., \citealp{Mirouh2019}, \HAS ). \citet{Bowman2018} and \citet{Barcelo2018} found a positive correlation between the effective temperature of \dsct stars and the frequency with the highest amplitude. According to \citet{2023ApJ...946L..10B}, the properties of pulsation spectra for \dsct stars display significant variation and do not exhibit a strong correlation with stellar temperature. These findings challenge the usefulness of the $\nu_{\mathrm{max}}$ relation for \dsct, at least for young stars.

In this paper, we present the observations and identification of \dsct stars observed by TESS cycle 4 in Section~\ref{section: TESS}. Then, we discuss the pulsation properties of these stars in Section~\ref{section: Chara}, and the period-luminosity relation of \dsct stars is investigated in Section~\ref{section: Period-Luminosity}. Section~\ref{section: Asteroseismic analysis} contains the asteroseismic analysis. Finally, Section~\ref{section: conclusions} presents the conclusions.

\section{Observations and target identification}\label{section: TESS}
The TESS satellite is equipped with four wide-angle cameras, each of which has a 24$^{\circ}\times96^{\circ}$ field of view, and each camera consists of four 2K$\times$2K CCDs \citep{Ricker2015}. The relatively low angular resolution of 21" per pixel is due to the large field of view. Additionally, each sector of observation is defined as the region of the sky observed over approximately 27.4 days, that is, two consecutive TESS orbits. After a sector is observed, TESS moves the field of view by 27$^{\circ}$ along the ecliptic to observe the next sector. TESS completes a full hemisphere scan in a single "cycle". The primary goal of the TESS mission is to detect planetary transits in bright stars across the celestial sphere. Specifically chosen stars undergo high-cadence observations, providing a valuable dataset for exploring previously uncharted astrophysical phenomena within main-sequence A and F stars. These data are particularly valuable for discovering new variability among stars.

\begin{figure}[!ht]
   \centering
   \includegraphics[width=\columnwidth]{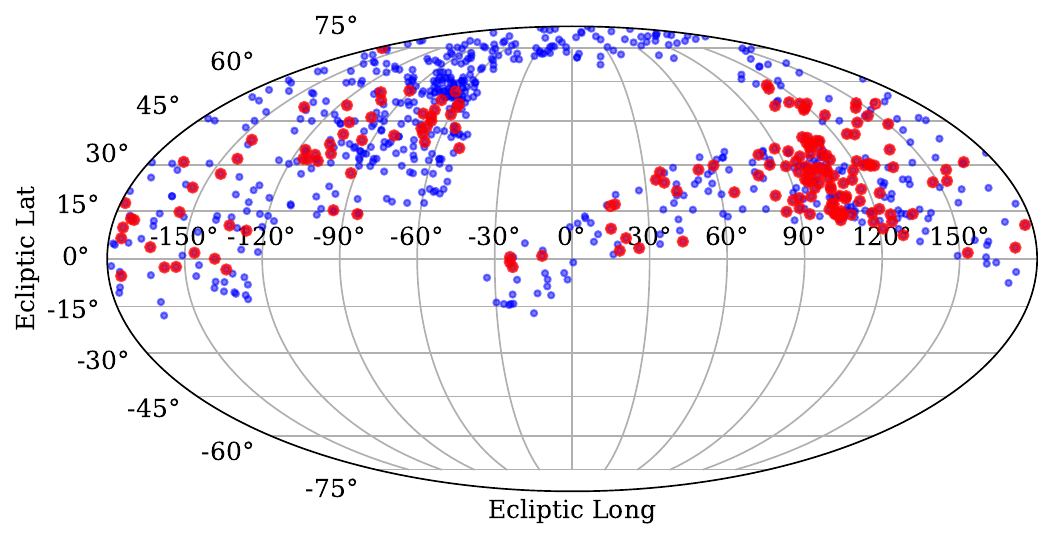}
   \caption{Position (Ecliptic latitude and longitude) of 765 \dsct stars (blue dots) and 179 \dsct stars with low-resolution spectral parameters (red dots) in the celestial sphere.}
              \label{Position}
    \end{figure}

We primarily used photometric data processed by the TESS Science Processing Operations Center \citep[SPOC,][]{Jenkins2016}, which is available at the Mikulski Archive for Space Telescopes (MAST)\footnote{https://archive.stsci.edu/}. The photometric data include Simple Aperture Photometry (SAP) and Pre-Search Data Conditioning (PDC) SAP flux, the latter eliminating long-term trends \citep{Twicken2010}. In order to detect new \dsct stars, we initially chose stars in cycle 4 within the typical temperature range of 6300K $\leq$ T$_{\rm eff}$ $\leq$ 9300K. We subsequently used the widely used, open-source LIGHTKURVE package \citep{Lightkurve2018} to compute a standard Lomb-Scargle periodogram to determine the oscillation frequencies of \dsct stars. We chose the frequency with the largest amplitude extracted in the frequency range studied as the main pulsation period and amplitude
of the star. For this analysis, frequencies with peaks in the range of $3 \leq \nu \leq 80$~day$^{-1}$ and with amplitude greater than 0.1 mmag were selected \citep{Bowman2016}. The Nyquist frequency of 2 min cadence observations is f$_{\rm N}$ = 360~day$^{-1}$, which is well above our limit range for the extracted frequencies. Following these criteria, 1056 potential stars were identified. Subsequently, we cross-referenced these stars with the existing \dsct star catalog \citep{Zhou2023}, which is a thorough collation of 42 previously published \dsct catalogs. To accurately identify the largest amplitude frequencies in the amplitude spectra, and to obtain the corresponding frequency errors. We utilized the FELIX tool (e.g., \citealp{Charpinet2010,Zong2016}). The uncertainty in frequency is determined using the method proposed by \citet{1999DSSN...13...28M}, allowing us to estimate the accuracy of our frequency measurements. Finally, we obtained 765 \dsct stars, which are listed in Table~\ref{tab:Catalog765}. The stellar parameters in the table are sourced from \citet{Stassun2019}; however, some parameters for certain stars were not provided. Of these stars, 179 have low-resolution spectral parameters from LAMOST \footnote{http://lamost.org/} \citep{Cui2012}, which are listed in Table~\ref{tab:Catalog179}. We directly utilized the parameters and errors obtained from LAMOST. Fig.~\ref{Position} shows the positions of all 765 \dsct stars discovered in TESS cycle-4.

\begin{figure}
        \centering
        \includegraphics[width=\columnwidth]{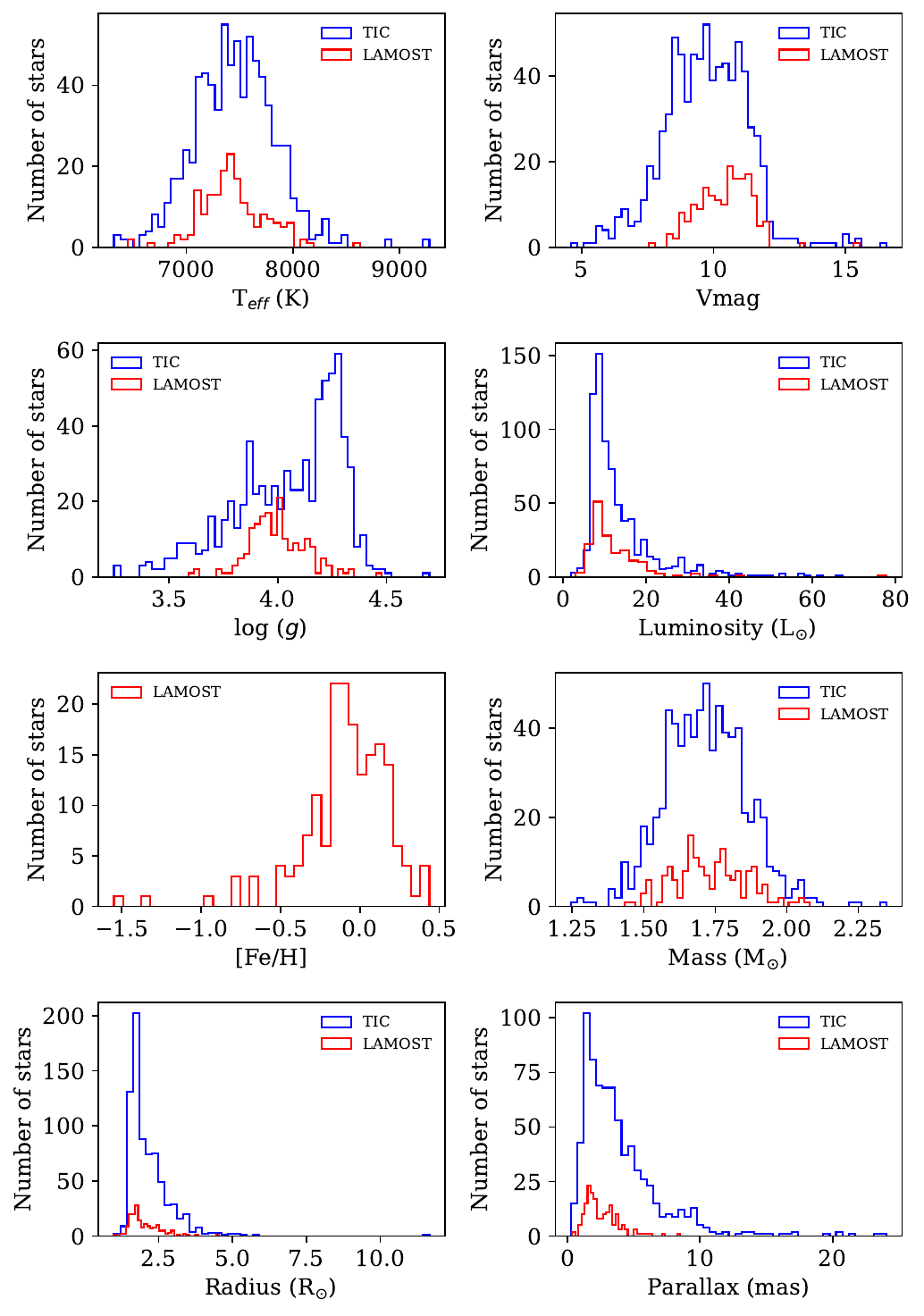}
        \caption{Effective temperature, $V$mag, surface gravity, luminosity, [Fe/H], mass, radius, and parallax distributions are shown for 765  and 179 \dsct stars as a blue and a red region, respectively.}
        \label{figure: parameters}
        \end{figure}

Figure~\ref{figure: parameters} shows the distribution of diverse parameters, including effective temperature, surface gravity, $V$mag, luminosity, [Fe/H], mass, radius, and parallax, for both the 765-star and 179-star samples. The mass, radius, effective temperature, and surface gravity values of \dsct stars lie within the ranges of 1.4M$_{\odot}$ to 2.1M$_{\odot}$, 1.2R$_{\odot}$ to 3.7R$_{\odot}$, 6300K to 9300K, and 3.3dex to 4.5dex, respectively. Examining the figure, it is evident that the two datasets largely maintain a consistent distribution across eight parameters. Notably, the surface gravity exhibits two distinct peaks in the TIC sample, whereas the LAMOST sample lacks the peak at surface gravity $\log g =$ 4.25. It appears that LAMOST performs more effectively with pre-main-sequence, late-main-sequence, and subgiant \dsct stars than with early-main-sequence \dsct stars. This discrepancy likely arises due to the smaller sample size of LAMOST. The median temperature hovers around 7500K, with median values of 1.75 for mass and 2 for radius, aligning with prior statistical findings for \dsct stars. $V$mag remains consistently close to ten, with only a handful of observed targets exceeding 12 mag, owing to the observational limitations of TESS.

\section{Pulsation properties of \dsct stars observed by TESS}\label{section: Chara}

According to the results of \citet{Dupret2005}, zero-age-main-sequence (ZAMS) stars located near the blue edge of the classical instability strip may exhibit higher order pulsations in the radial overtone mode compared to stars located at the red edge, leading to higher observed pulsation frequencies. This is caused by the different depths of the He~{\sc ii} driving region in the stellar envelope about T$_{\rm eff}$;  if the star is too hot, the ionization region will be located near the surface where the fluid is too tenuous to drive the oscillations, and if the star is too cold, convection will suppress the oscillations.

\begin{figure}[htpb]
                \centering
                \includegraphics[width=\columnwidth]{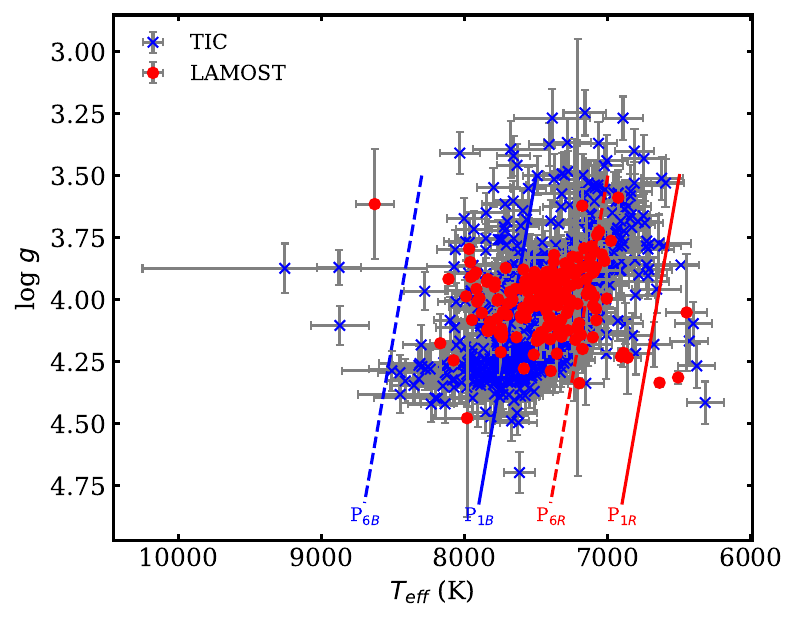}
        \caption{T$_{\rm eff} - \log g$ diagram for our 765 \dsct stars observed by TESS. The blue dots represent the positions of \dsct stars using parameters from the TIC, while the red dots correspond to a subgroup of \dsct stars with spectroscopic T$_{\rm eff}$ and $\log g$ values. The solid blue and red lines indicate the theoretical blue and red edges of the classical instability strip, respectively, for p~modes with radial order $n=1$, while the dashed blue and red lines represent radial order $n=6$ according to \citet{Dupret2005}.}
        \label{figure: T_logg}
        \end{figure}

Figure~\ref{figure: T_logg} shows the T$_{\rm eff} - \log g$ diagrams for our 765 \dsct stars. The blue and red dots correspond to T$_{\rm eff}$ and $\log g$ obtained by TIC and LAMOST, respectively. The blue and red lines in the diagram represent the theoretical blue and red edges of the classical instability band, respectively; the solid lines are for p~modes of radial order $n=1$ and the dashed lines are for radial order $n=6$ \citep{Bowman2018}. Most of the \dsct stars are located in the instability strip, and the subgroup of stars with LAMOST spectral parameters seems to be more tightly clustered, suggesting that the spectral parameters may provide more accurate estimates of T$_{\rm eff}$ and $\log g$ for these \dsct stars. The number of stars distributed outside the classical instability strip is likely to be the result of inaccurate parameters or the presence of modes at a higher radial order.

        \begin{figure}[htp]
        \centering
        \includegraphics[width=\columnwidth]{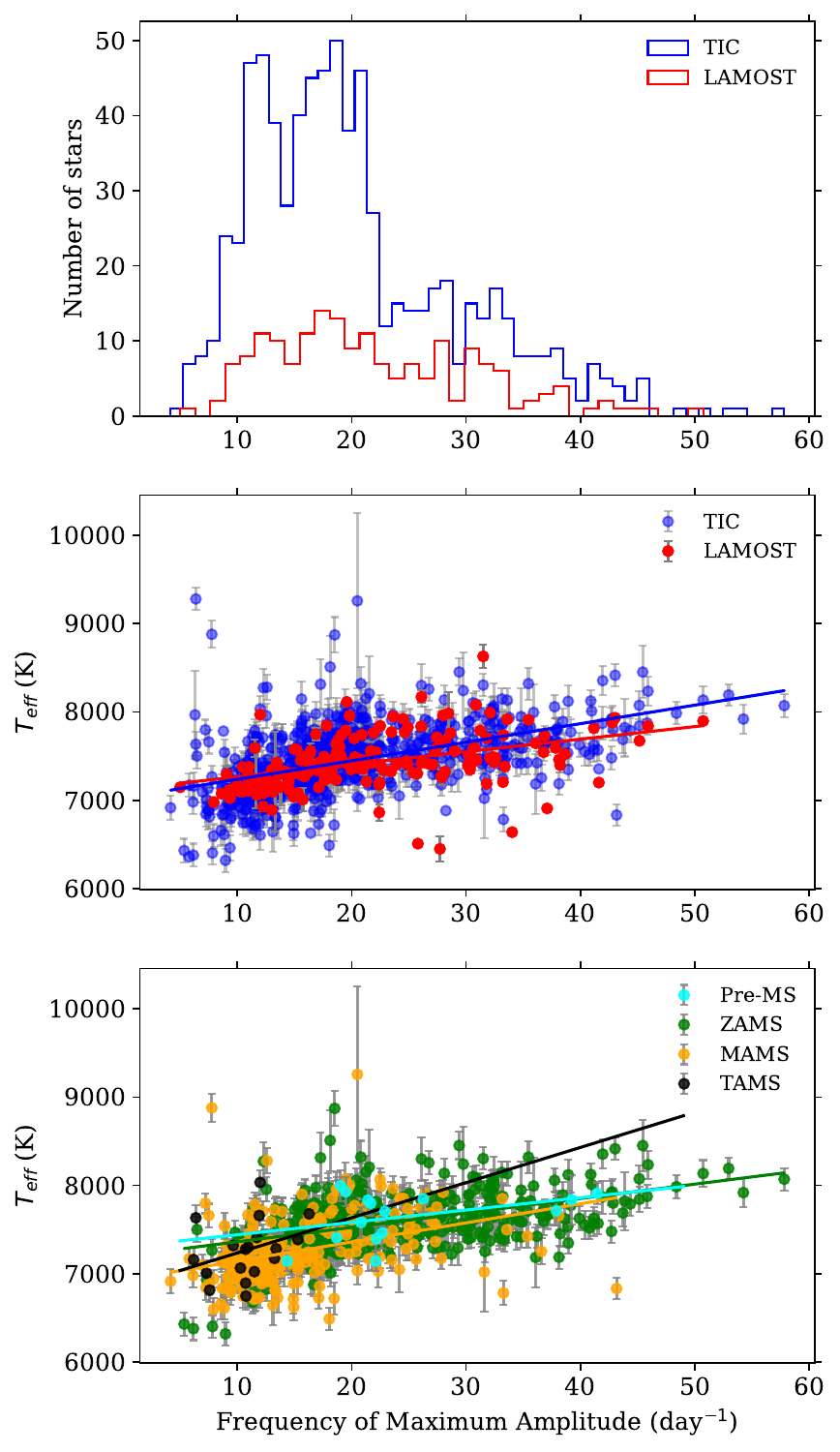}
        \caption{The relationship
between pulsation frequency and temperature was investigated
for both TIC and LAMOST samples. Top row shows distribution of frequency with maximum amplitude (shown as a blue region) for the ensemble of our 765 \dsct stars observed by TESS. The red region indicates the distribution for the subgroup of stars with spectroscopic parameters from LAMOST. The middle row shows the frequency of the maximum amplitude plotted against the effective temperature, with the TIC and LAMOST parameters represented by blue and red dots, respectively. The statistically significant linear regressions are displayed as solid lines. The bottom row shows the same relationship as the middle row, dividing the stars into four groups of TIC stars according to surface gravity with $\log g \geq 4.0$ representing ZAMS, $3.5 \leq \log g < 4.0$ representing the mid-age main sequence (MAMS), and $\log g < 3.5$ representing the terminal age main sequence (TAMS), as well as Pre-MS stars matched to the literature.}
        \label{figure: fre_teff}
        \end{figure}

\begin{table}
        \tiny
        \centering
	\caption{The statistics of linear regressions of the frequency of maximum amplitude and effective temperature for the \dsct stars in this study, using both TIC values and spectroscopic values from LAMOST. The regressions are segmented based on the log $g$ subgroups present in the TIC data and Pre-MS \dsct stars}.

	\begin{tabular}{l r r r r r }
	\hline
    \hline
	& \multicolumn{1}{c}{$N_{\rm stars}$} & \multicolumn{1}{c}{$m$} & \multicolumn{1}{c}{$c$} & \multicolumn{1}{c}{$R$} & \multicolumn{1}{c}{$p$-value} \\
        & \multicolumn{1}{c}{} & \multicolumn{1}{c}{(K / day$^{-1}$)} & \multicolumn{1}{c}{(K)} & \multicolumn{1}{c}{} & \multicolumn{1}{c}{}\\
	\hline
	\multicolumn{1}{c}{TIC} \\

	All			&	{765}	&	{$20.9 \pm 2.6$}	&	{$7026.5 \pm 0.2$}	&	{0.50}	&	{$< 0.0001$}\\

        Pre-MS      &	{15}	&	{$14.0 \pm 18.6$}	&	{$7301.8 \pm 15.8$}	&	{0.41}	&	{$0.1279$}\\

	ZAMS 		&	{434}	&	{$16.3 \pm 2.9$}	&	{$7198.7 \pm 0.4$}	&	{0.47}	&	{$< 0.0001$}\\

	MAMS	    &	{262}	&	{$21.6 \pm 6.8$}	&	{$6927.6 \pm 1.1$}	&	{0.36}	&	{$< 0.0001$}\\

	TAMS		&	{18}	&	{$39.8 \pm 59.7$}	&	{$6837.1 \pm 58.4$}	&	{0.33}	&	{$0.1769$}\\

	\hline
	\multicolumn{1}{c}{LAMOST} \\

	All 	    &	{179}	&	{$14.2 \pm 4.6$}	&	{$7122.1 \pm 0.9$}	&	{0.42}	&	{$< 0.0001$}\\

	\hline
	\end{tabular}
	\label{table: fit stats}
	\end{table}

\citet{Bowman2018} analyzed the statistics of the stellar parameters and pulsation modes of \dsct stars from $Kepler$ observations, and we performed a similar operation for the first time for \dsct stars from TESS observations. The relationship between pulsation frequency and temperature was investigated for both TIC and LAMOST samples. The top row of Fig.\ref{figure: fre_teff} shows the distribution of the frequency with the maximum amplitude of 765 \dsct stars. The blue region represents the distribution with TIC parameters, while the red region shows the distribution for a subgroup of stars with LAMOST parameters for comparison. No \dsct star in our sample show maximum amplitude pulsations at frequencies higher than 60~days$^{-1}$, which confirms the results of the statistical analysis performed by \citet{Bowman2018}. The frequency with the maximum amplitude of \dsct stars lies in the range of $5 < \nu \leq 60$~day$^{-1}$ for TESS observations.

The frequency of maximum amplitude plotted against the effective temperature for the two samples of TIC and LAMOST are indicated by blue and red dots in the middle row of Fig.\ref{figure: fre_teff}, respectively, and the blue and red lines represent statistically significant linear regressions for the two samples. Linear regressions of the frequency of maximum amplitude and T$_{\rm eff}$ using either the TIC or LAMOST values exhibit statistically significant positive correlations, with R = 0.50 and R = 0.42, respectively. It is worth noting that the slope of the LAMOST fit is smaller than that of the TIC fit. The gradient, $m$, intercept, $c$, coefficient of correlation, $R$, and $p$-value for each data set are listed in Table \ref{table: fit stats}.

During the main-sequence phase, as the stellar radius increases, the pulsation frequency of a star's modes decreases, but this decrease occurs at a relatively slow rate \citep{ASTERO}. Therefore, in addition to the effective temperature, the pulsation frequency of a star is also influenced by its surface gravity. Both T$_{\rm eff}$ and $\log g$ play important roles in determining the pulsation frequency of a star, indicating that the frequency is dependent on both temperature and the stellar structure, as represented by the surface gravity. Thus, we investigate the relationship between pulsation frequency and T$_{\rm eff}$ by classifying the stars of the TIC sample according to different $\log g$ values similarly to \citet{Bowman2018}. In comparison to \citet{Bowman2018}, we expanded our Pre-MS \dsct star sample by cross-referencing our sample with more than ten catalogs, resulting in the identification of 15 Pre-MS \dsct stars. The remaining TIC stars are grouped according to $\log g \geq 4.0$ (ZAMS), $3.5 \leq \log g < 4.0$ (MAMS), and $\log g < 3.5$ (terminal-age main sequence; TAMS). Undocumented Pre-MS \dsct stars might be confused with MAMS stars due to their similar $\log g$ values \citep{Steindl2021}, even if the relative durations of these phases make Pre-MS stars relatively rare. In the bottom panel of Fig.\ref{figure: fre_teff}, we show Pre-MS, ZAMS, MAMS, and TAMS, which are shown as cyan, green, yellow, and black dots, respectively. Lines of the same color as the sample dots indicate the respective linear regressions. We did not perform the same classification for the 179 \dsct stars with LAMOST parameters due to the small number of stars with them. The linear regression results are listed in Table\ref{table: fit stats}. According to our statistical results, the T$_{\rm eff}$ and maximum amplitude pulsation frequency of these four different groups present different fitting results. The R values for ZAMS \dsct stars and MAMS \dsct stars are 0.47 and 0.36, respectively. Although the values are not significant, the $p$ values are quite small; this statistic is still meaningful, and the relationship between temperature and maximum amplitude pulsation frequency is more significant for MAMS stars than for ZAMS stars. Since there are too few stars in the Pre-MS and TAMS, reliable statistical results cannot be obtained ($p$ = 0.1279 and $p$ = 0.1769), according to the trends reflected in ZAMS and MAMS, the relationship between the temperature and the maximum amplitude pulsation frequency of TAMS stars should be more significant. A follow-up analysis of Pre-MS and TAMS \dsct stars will be necessary to obtain accurate results.

\section{Period–luminosity relation for \dsct stars}
\label{section: Period-Luminosity}
        \begin{figure}
        \centering
        \includegraphics[width=\columnwidth]{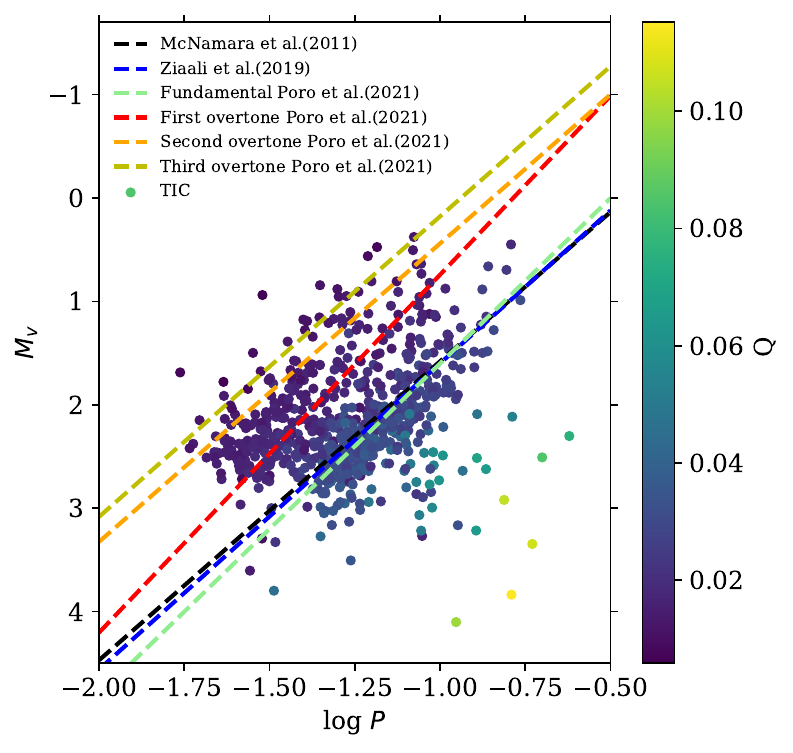}
        \includegraphics[width=\columnwidth]{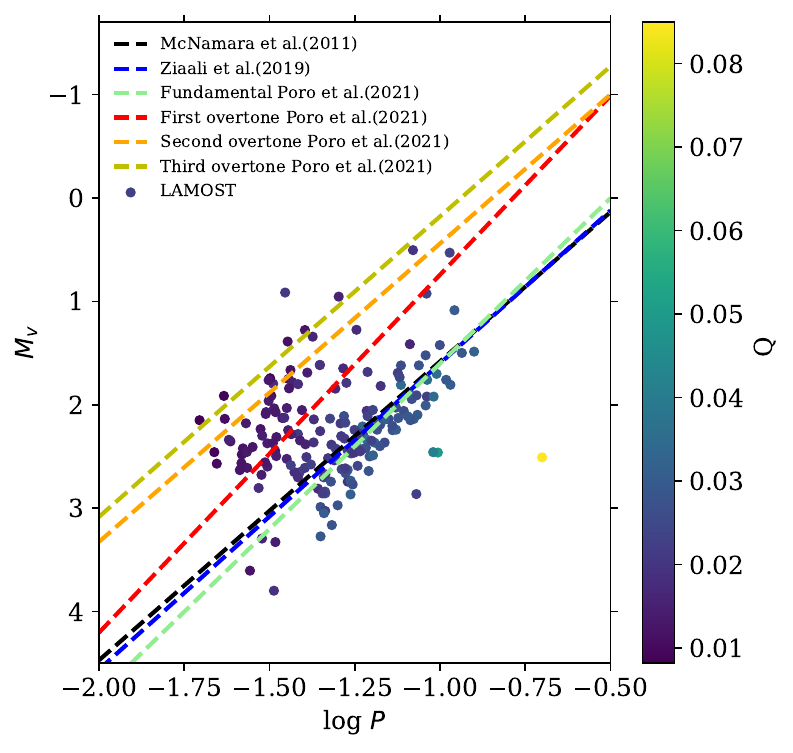}
        \caption{The P-L relation
of TIC and LAMOST samples, respectively. Upper row shows P-L relation of TIC, and the bottom row shows the P-L relation of LAMOST. The black dashed line is the P-L relation for high-amplitude \dsct stars derived by \citet{2011AJ....142..110M}, while the P-L relation obtained by \citet{2019MNRAS.486.4348Z} is shown by the blue dashed line. The remaining dashed lines represent different modes of the P-L relation, as derived by \citet{2021PASP..133h4201P}. Color-coding represents the different values of the pulsation constant, $Q$. }
        \label{figure: PLR}
        \end{figure}

The relations between pulsation periods and the luminosity of pulsating stars, known as period-luminosity (P-L) relations, have been studied extensively and have a rich history dating back to \citet{Leavitt1912}. When a group of pulsating stars falls within a relatively limited range of T$_{\rm eff}$, their luminosity becomes highly correlated with their stellar density and, as a result, with the periods of their pressure modes. This phenomenon has been observed and described by scientists (e.g., \citealp{Eddington1926,Carroll2006}). To obtain the P-L relations for the two samples, TIC and LAMOST, we performed a series of calculations on the absolute magnitude $M_v$ and some other parameters. The extinction coefficient $A_v$ was determined based on the following equation:
\begin{equation}
A_{v}=R_{v} \times E(B-V),
\label{equation1}
\end{equation}
where the coefficient $R_v = 3.1$ \citep{1999PASP..111...63F}. The equation below was used to calculate $M_v$:
\begin{equation}
M_{v}=V - 5\log d + 5 - A_{v},
\label{equation1}
\end{equation}
$M_v$ and $V$ refer to the absolute and apparent magnitudes of a star measured in the $V$ band. The distance ($d$) of the star in parsecs was determined using parallax measurements from Gaia DR3. Additionally, the bolometric absolute magnitude ($M_{bol}$) was calculated using the formula $M_{bol} = M_v + BC$, where $BC$ is the bolometric correction. The value of $BC$ was obtained by applying polynomial transformation equations developed by \citet{1996ApJ...469..355F} and corrected by \citet{2010AJ....140.1158T}. The P-L of different radial pulsation modes are distinct from each other, as noted in the study by \citet{2021PASP..133h4201P}. One way to distinguish between different pulsation modes is to calculate their pulsation constants using the formula $Q = P \sqrt{\frac{\overline{\rho}}{\overline{\rho}_{\odot}}}$, where $Q$ is the pulsation constant expressed in units of days, $P$ is the pulsation period also units of days, $\overline{\rho}$ is the average density of the star, and $\overline{\rho}_{\odot}$ is the average density of the Sun, which serves as a normalization factor. The pulsation constant is a measure of how the pulsation period of a star is related to its average density. The above equation can be rewritten as \citep{1990ASPC...11..263B}:
\begin{equation}
\log Q= \log P + \frac{1}{2}\log g + \frac{1}{10} M_{\mathrm{bol}} + T_{\mathrm{eff}} -6.454.
\label{equation1}
\end{equation}
We obtained the $Q$ value of each \dsct star and listed it in Table~\ref{tab:Catalog765}.

The upper and lower rows of Fig. \ref{figure: PLR} show the P-L relation of TIC and LAMOST, respectively. The P-L relation for high-amplitude \dsct stars derived by \citet{2011AJ....142..110M} is represented by the black dashed line, while the P-L relation obtained by \citet{2019MNRAS.486.4348Z} is shown by the blue dashed line. Additionally, various dashed lines in the figure represent different modes of the P-L relation, as derived by \citet{2021PASP..133h4201P}. The color-coding in Fig. \ref{figure: PLR} represents the different values of the pulsation constant, $Q$. Typically, the values of the pulsation constant for the fundamental, first, and second overtone radial p modes in \dsct stars fall within the range of $0.022 \leq Q \leq 0.033\ \text{d}$, as reported by \citet{1975ApJ...200..343B}. However, it should be noted that there can be significant uncertainties in the determination of pulsation constants, with fractional uncertainties reported by \citet{Breger1990} as high as 18 percent. This can lead to the misidentification of a first overtone radial mode as either the fundamental or second overtone radial mode, emphasizing the need for caution when applying this method of mode identification \citep{Bowman2018}. Therefore, determining the accurate pulsation order of each star can be challenging, and additional observations or modeling may be required to confirm the resulting mode identification. The stars with LAMOST parameters are basically all in good agreement with the P-L relation, which confirms that our judgment of this sample of stars is accurate. The availability of LAMOST spectra (and associated T$_{\rm eff}$, log $g$) also opens up the possibility of further analytical studies, and therefore our next step will be to analyze the asteroseismic indices of this fraction of stars. The outliers in the bottom right part of Fig. \ref{figure: PLR} are the stars lying on the red side of the instability strip or colder than the instability strip's red edge, and they are most likely misidentified $\gamma$ Dor stars.

    \begin{figure}[htpb!]
    \centering
    \includegraphics[width=\columnwidth]{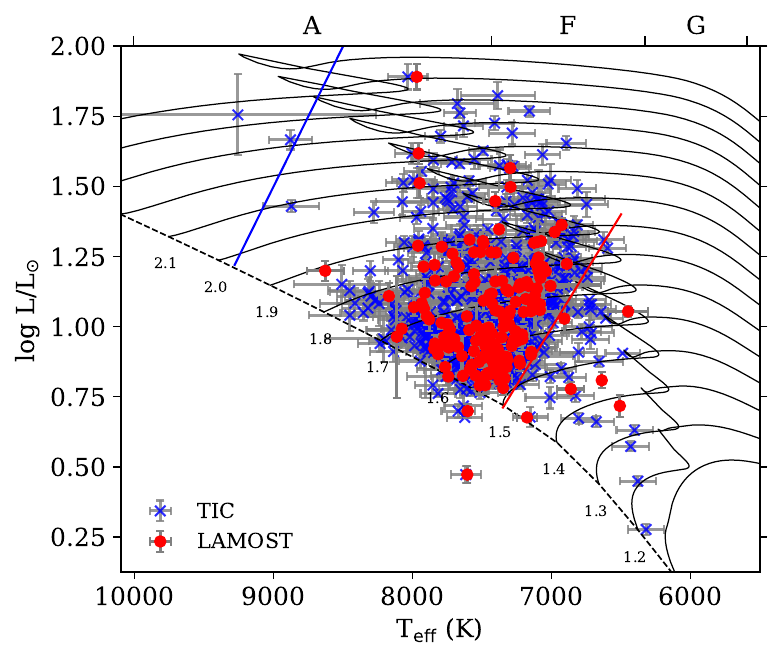}
    \caption{HR diagram of \dsct stars. The blue dots represent the location of the \dsct stars using the TIC parameters, while the red dots represent the subgroup of stars with spectroscopic values of T$_{\rm eff}$ from LAMOST. The observational instability strip boundaries from \citet{Murphy2019} are shown as solid lines. The evolutionary tracks were computed using MESA v10108 with X = 0.71 and Z = 0.01.}
    \label{figure: HRDtic_lamost}
    \end{figure}

Figure \ref{figure: HRDtic_lamost} shows the HR diagram of the \dsct stars. The blue dots indicate the positions of the \dsct stars using the TIC parameters, while the red dots represent the subgroup of stars with spectroscopic values of T$_{\rm eff}$ obtained from LAMOST. The evolutionary tracks were computed using MESA v10108 with X = 0.71 and Z = 0.01 \citep{Murphy2019}. The observational instability strip boundaries derived by \citet{Murphy2019} are shown as solid lines. Most of the \dsct stars are located within the pulsating instability strip.

\section{Asteroseismic analysis}
\label{section: Asteroseismic analysis}

\begin{figure}[ht!]
        \centering
        \includegraphics[width=\columnwidth]{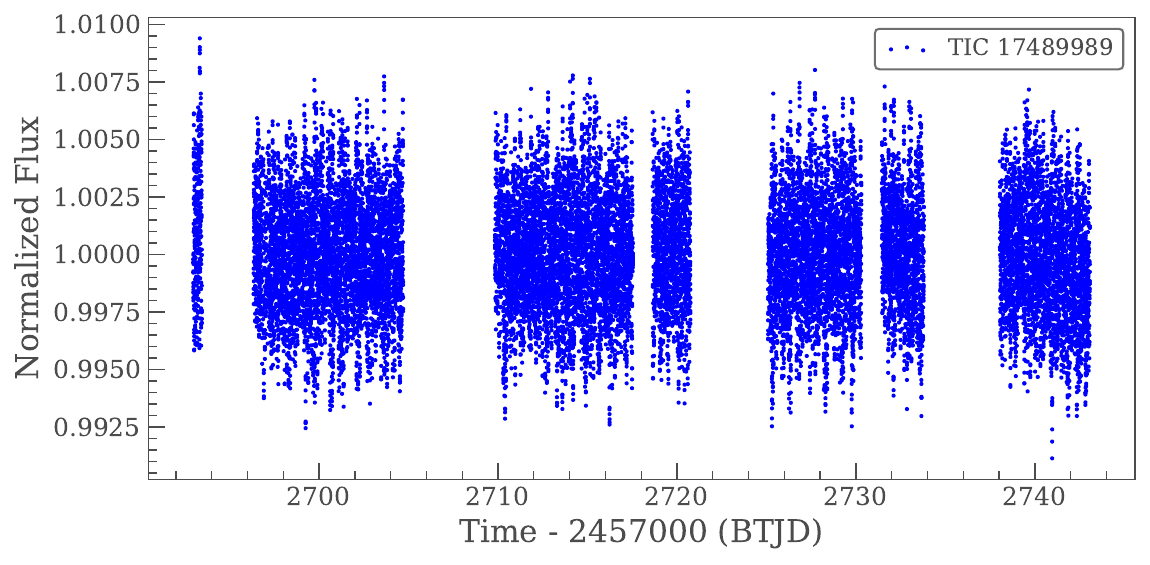}
        \caption{Normalized light curve of TIC 17489989 captured by TESS during sectors 51 $\&$ 52.}
        \label{figure: TIC17489989-lc}
        \end{figure}

\begin{figure}[ht!]
        \centering
        \includegraphics[width=\columnwidth]{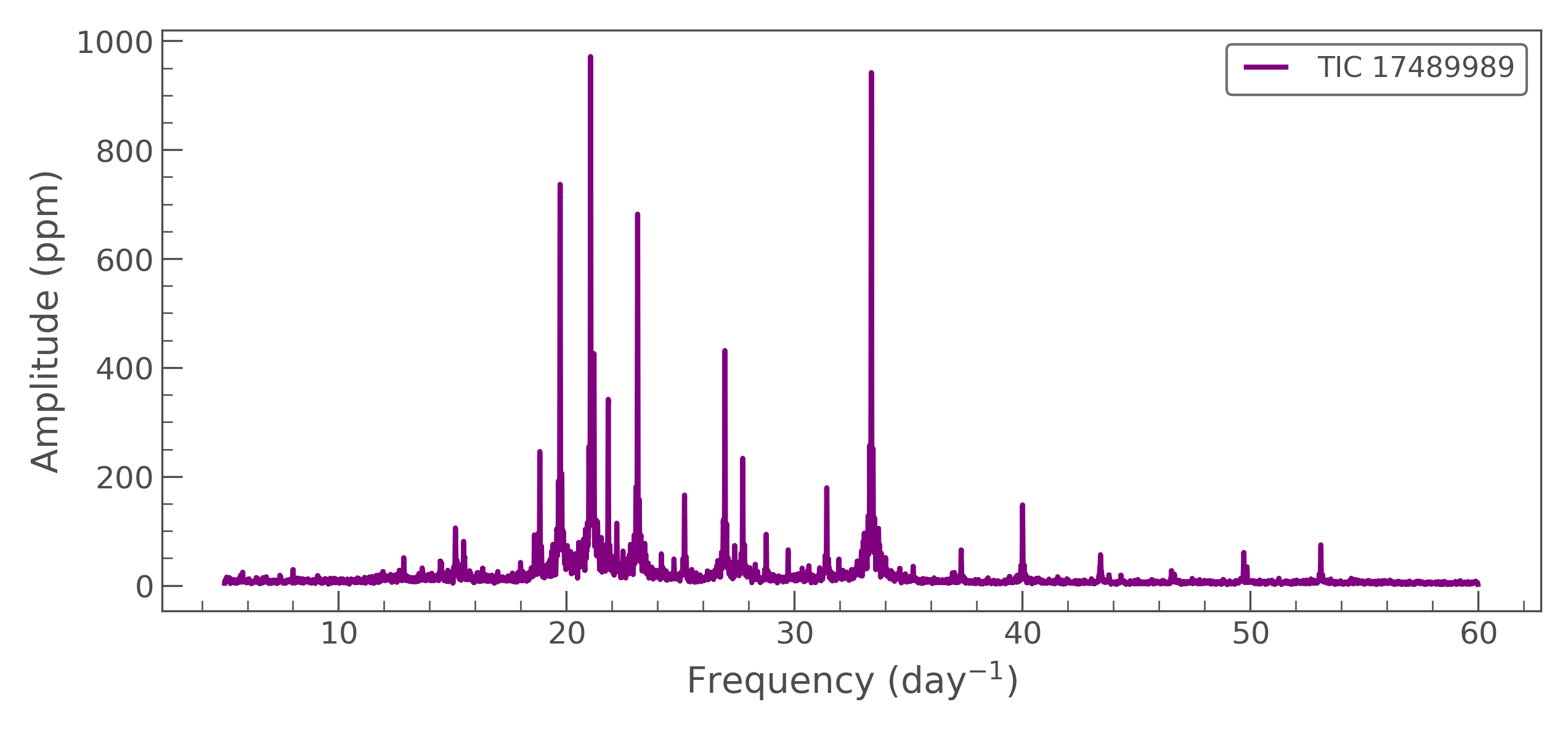}
        \caption{Lomb-Scargle periodogram depicts relationship between amplitude (measured in ppm) and frequency.}
        \label{figure: TIC17489989-power}
        \end{figure}

In order to analyze the TESS cycle-4 short-cadence light curves of 179 \dsct stars with LAMOST parameters, the Python LIGHTKURVE package was used to download and analyze TESS data. We eliminated 4$\sigma$ outlier data points from each star's light curve. Then, the outcome is smoothed \citep{Savitzky1964} and normalized to mean intensity, as shown in Fig.\ref{figure: TIC17489989-lc}, and the asteroseismic power spectrum was derived, as shown in Fig.\ref{figure: TIC17489989-power}. We only retained frequencies with a signal-to-noise ratio (SNR) of at least four \citep{Breger1993}.

\begin{figure}[ht!]
        \centering
        \includegraphics[width=\columnwidth]{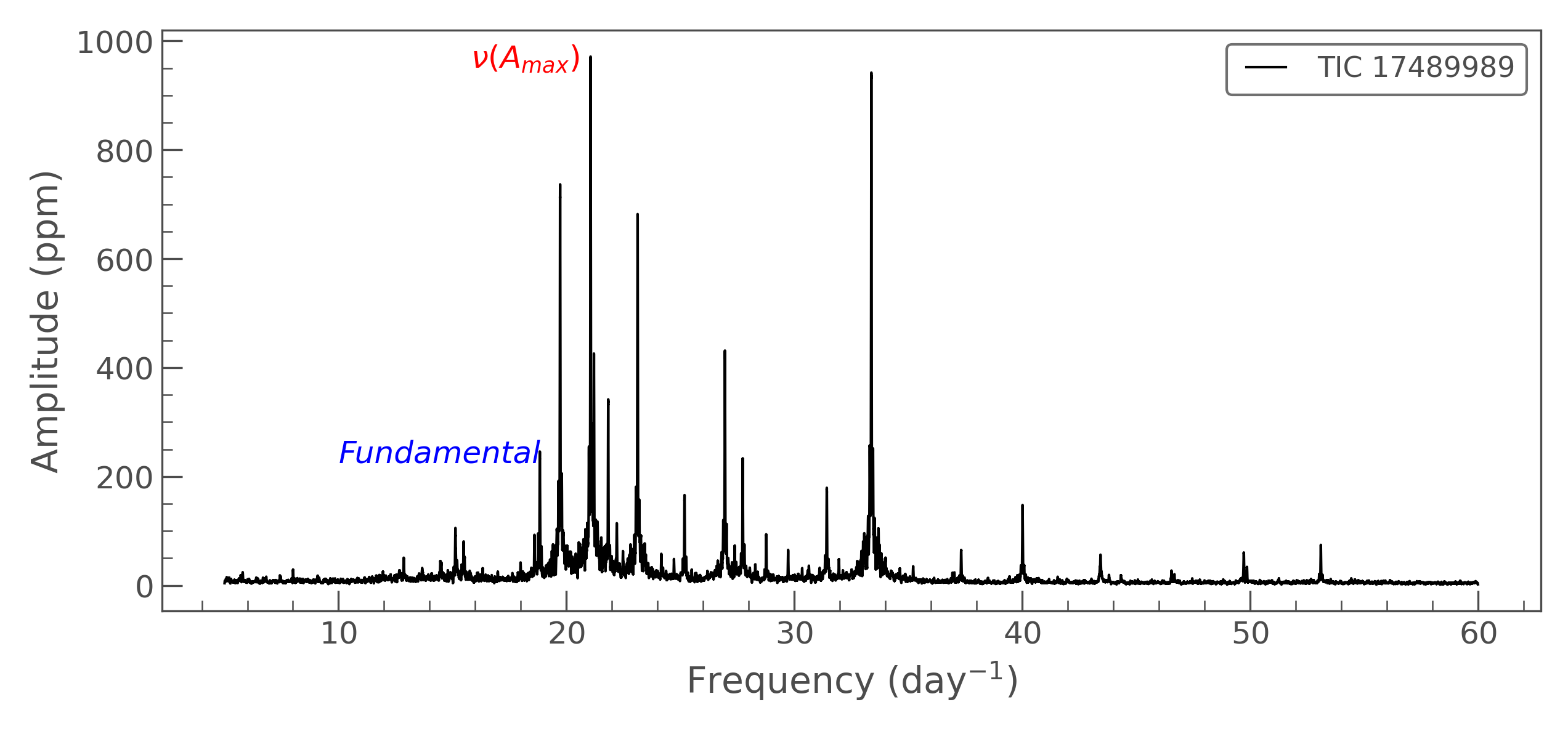}
        \caption{For TIC 17489989, the fundamental frequency and $\nu$(A${\rm max}$) are shown in blue and red, respectively.}
        \label{figure: TIC17489989_funadamental-maximum}
        \end{figure}

\begin{figure}[ht!]
        \centering
        \includegraphics[width=1.1\columnwidth]{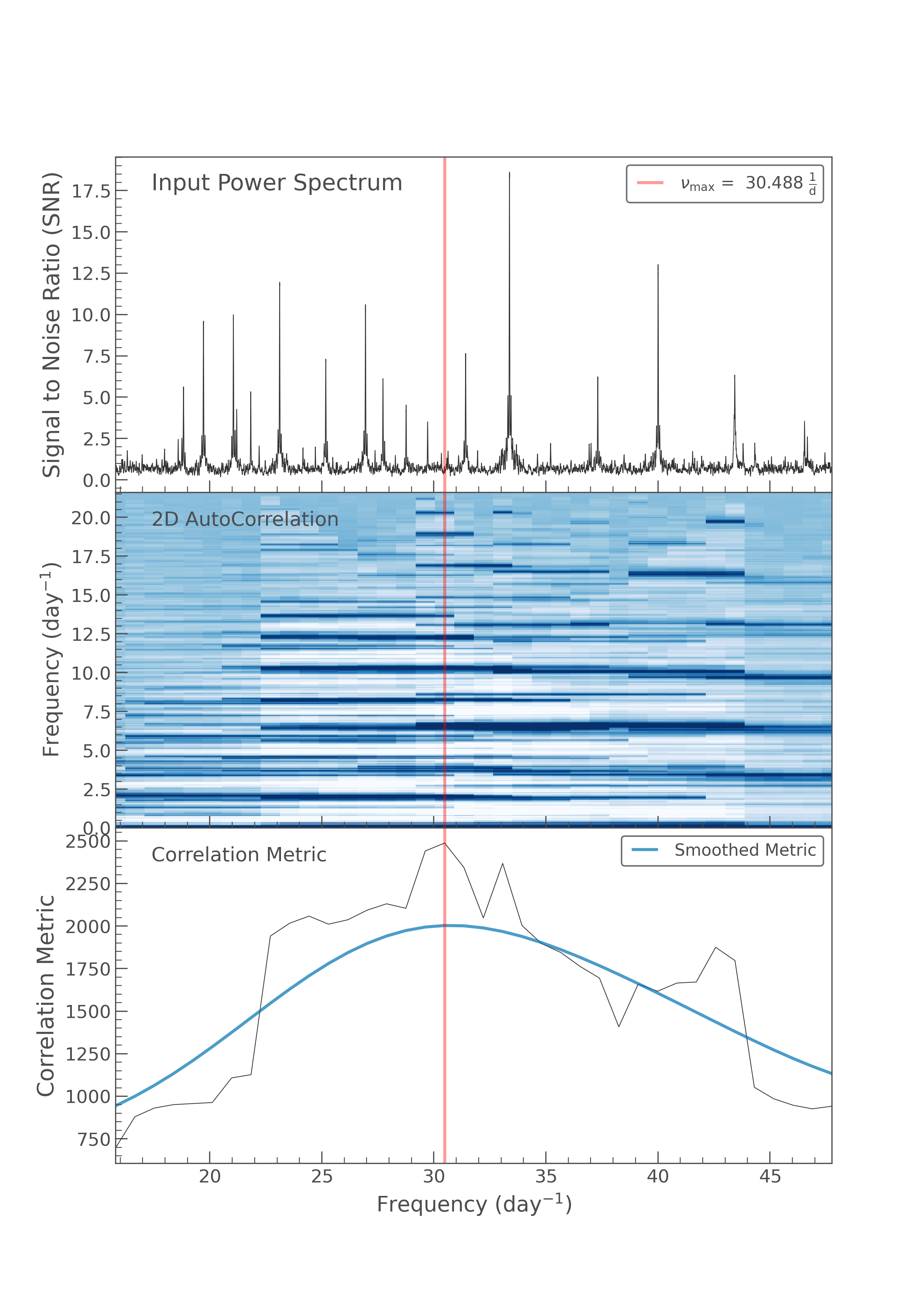}
        \caption{SNR frequency spectra for TIC 17489989 is in the upper panel. The $\nu_{\rm max}$ is indicated by the red line. Frequency spectra with 2D autocorrelation applied are in the middle panel. Lower panel: Gaussian-smoothed curve (blue line) and the mean collapsed correlation shown against the frequency \citep[see][]{Viani2019}.}
        \label{figure: TIC17489989-numax}
        \end{figure}

The fundamental radial mode was typically thought to be the frequency with the maximum amplitude, but earlier investigations of \dsct stars (e.g., \citealp{Murphy2020,2020Natur.581..147B}, \HAS ) revealed that, in fact, it is not always the largest peak in the power spectrum. We now describe our combined criteria to identify the fundamental mode. We consider the lowest frequency in the power spectrum with an amplitude greater than 0.2 times the highest peak as the fundamental radial mode (\HAS ).
The frequency associated with the highest amplitude in each star's power spectrum is called $\nu({\rm Amax})$. We picked TIC 17489989 as an example and used its data to illustrate the analyses. Fig.\ref{figure: TIC17489989_funadamental-maximum} shows the fundamental frequency and $\nu({\rm Amax}$) of TIC 17489989 in blue and red, respectively. \HAS~ suggested using the frequency of the envelope's peak $\nu_{\rm max}$ rather than $\nu({\rm Amax})$ in an SNR frequency spectra. The 2D autocorrelation technique applied to the frequency spectra to obtain $\nu_{\rm max}$ is shown in Fig.\ref{figure: TIC17489989-numax}. This method was described in detail by \citet{Viani2019} and has been applied to \dsct stars in \HAS\ and \citet{Pamos2022}.

    \begin{figure}[ht!]
        \centering
        \includegraphics[width=1.1\columnwidth]{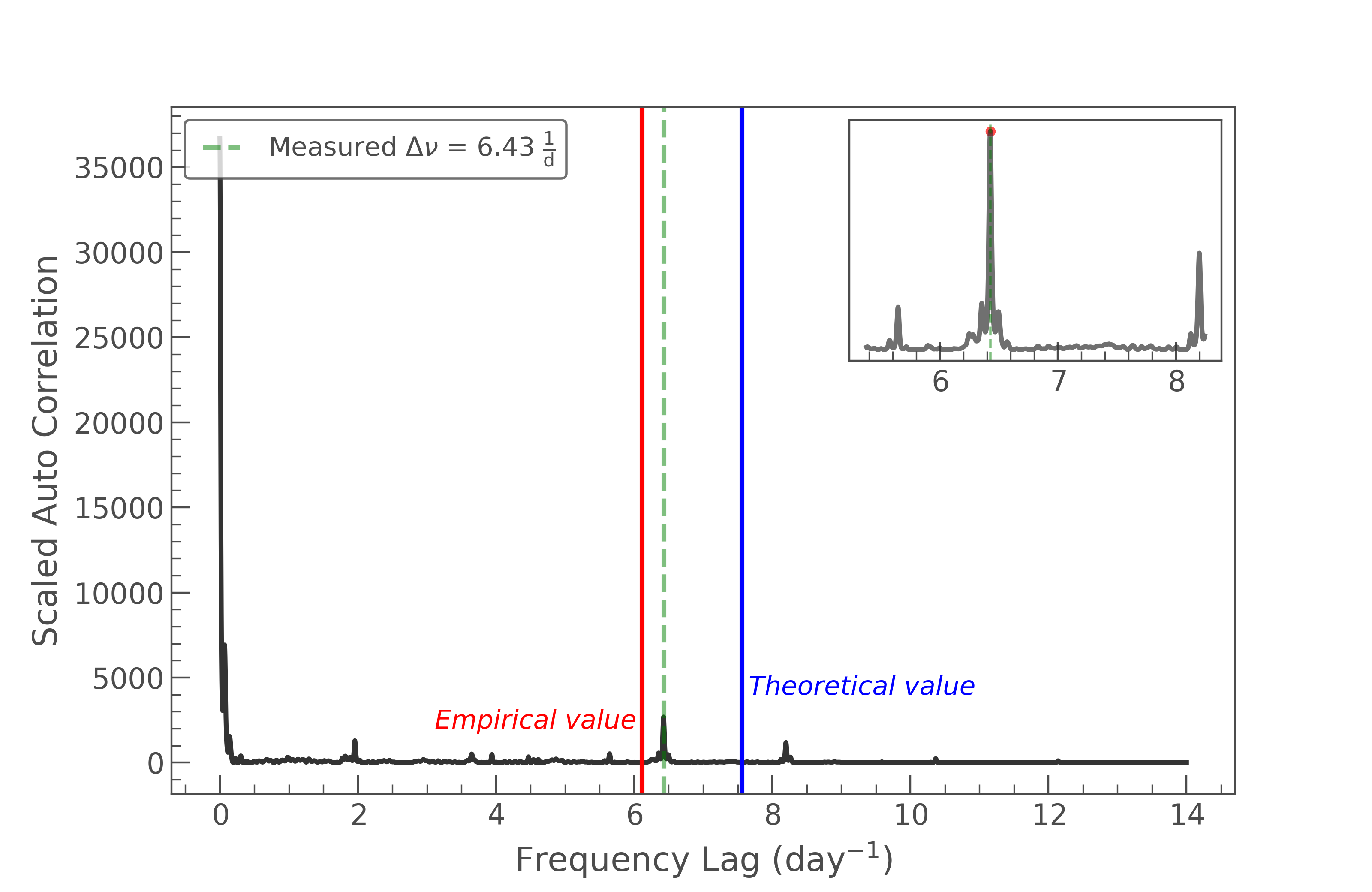}
        \caption{2D autocorrelation was used to determine large frequency separation $\Delta\nu$ of TIC 17489989. Blue, red, and dashed green lines, respectively, show the theoretical, empirical, and obtained values.}
        \label{figure: TIC17489989-deltanu}
        \end{figure}

     \begin{figure}[ht!]
        \centering
        \includegraphics[width=1.1\columnwidth]{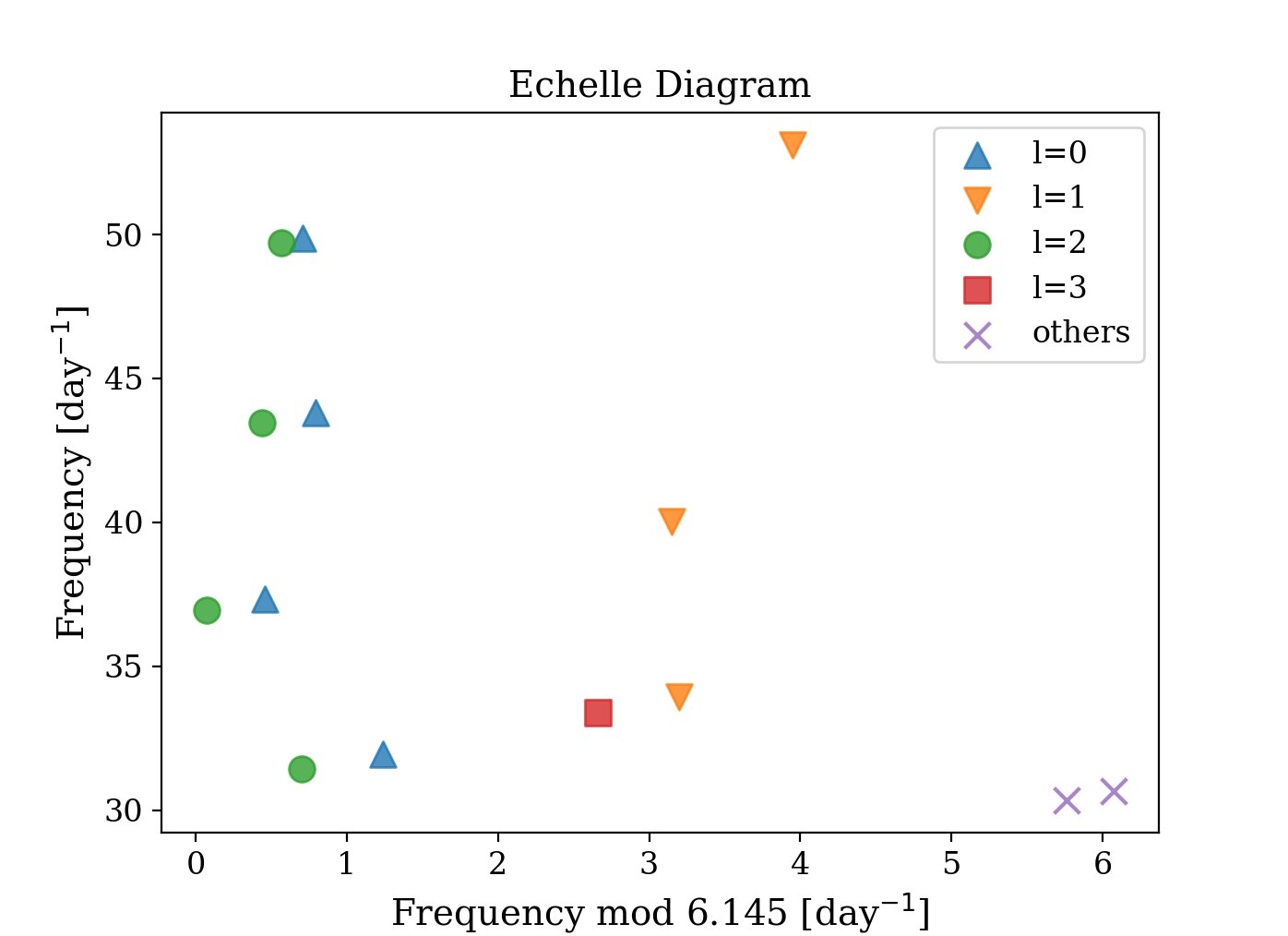}
        \caption{Echelle diagram for TIC 17489989 with large frequency separation of approximately 6.145 day$^{-1}$. Different colored points represent different pulsation modes.}
        \label{figure: TIC17489989-echelle_mark}
        \end{figure}

The large frequency separation $\Delta\nu$, that is, the average frequency interval between modes at successive radial orders $n$ and the angular degree $l$, is a crucial index.
In order to calculate $\Delta\nu$, we took into account a window centered on $\nu_{\rm max}$ (approximately containing all of the mode peaks) for each SNR frequency spectra. Then, the $\Delta\nu$ was determined through the 2D autocorrelation method as shown in Fig.\ref{figure: TIC17489989-deltanu}. We estimated the final value for the large frequency separation by employing an automatic peak finding process (e.g., \citealp{Huber2009,Mosser2009}) for the peaks that are around an empirical value (based on the relationship between the density and $\Delta\nu$). A theoretical relationship between the mean stellar density ($\rho$) and the large separation $\Delta\nu$ of high-order modes (in the asymptotic regime) yields $\Delta\nu \propto\sqrt{\rho}$.
Many works have investigated the empirical relationship between the mean density and for \dsct stars. According to \citet{suarez2014}, $\Delta\nu$ and mean density could be expressed empirically as $\Delta\nu \propto 0.78\rho^{0.46}$. \citet{Garc2015, Garc2017} used accurate parameters of eclipsing binaries to constrain the relation to $\Delta\nu \propto 0.82\rho^{0.49}$, while \citet{2020Natur.581..147B} found a relation, $\Delta\nu \propto 0.85\sqrt{\rho}$ , for their sample of young stars. Based on two years of TESS data for 438 \dsct stars, \HAS\  modified the relation between the large frequency separation and stellar density as $\Delta\nu \propto 0.76\rho^{0.43}$. In our analysis, we took into account the large frequency separation of less than 35\% divergence from the empirical scaling relation of \HAS . As a result, 148 targets remain on our final list and are listed in Table~\ref{tab:Catalog179}.

    \begin{figure}[ht!]
        \centering
        \includegraphics[width=0.97\columnwidth]{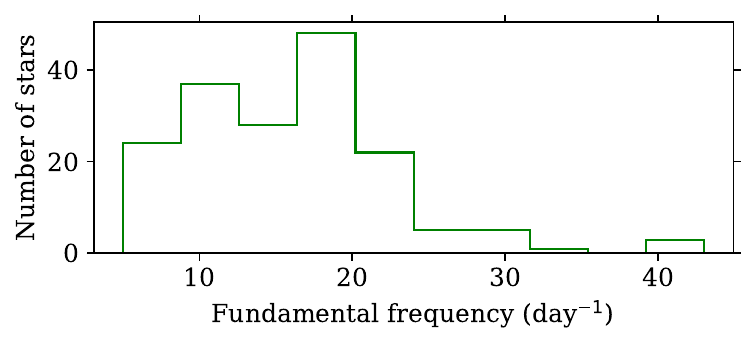}
        \includegraphics[width=0.97\columnwidth]{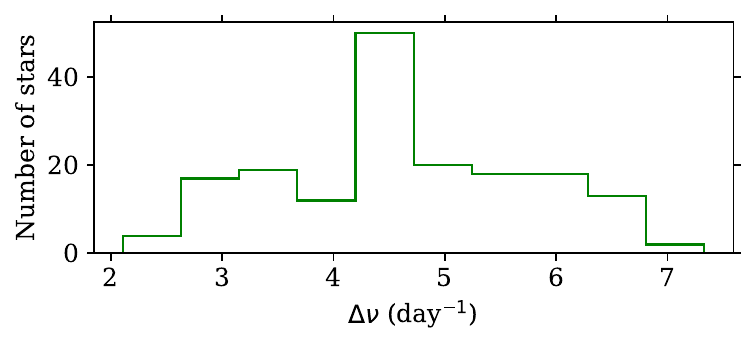}
        \caption{Frequency distribution of fundamental mode (day$^{-1}$) and large separation (day$^{-1}$) for 148 \dsct stars.}
        \label{figure: fundament}
        \end{figure}

Some of the targets show regular frequency spacing in the \'echelle diagram. Fig.\ref{figure: TIC17489989-echelle_mark} shows the \'echelle diagram of TIC 17489989. We fit a Gaussian envelope to the SNR periodogram and took the peak of the Gaussian as the central frequency to search for a pattern around. However, there is no gaussian envelope in this case. To obtain a better $\Delta \nu$, we tried different frequency intervals to search for $\Delta \nu,$ and it seems that the ridges appear for frequencies above 30 days$^{-1}$. Not all the frequencies are present, though. The result of the manual \footnote{https://github.com/javier-iaa/echelle\_slider} search provides a more accurate result than the autocorrelation method, which is a rough approximation based on the asymptotic regime for solar stars. It is, however, a method that can be implemented in an automatic algorithm, so we could cautiously consider the results a rough approximation to the large frequency separation. For the particular case shown in Fig.\ref{figure: TIC17489989-echelle_mark}, the star TIC 17489989 shows a $\Delta \nu$ of 6.145 day$^{-1}$, while it was 6.430 days$^{-1}$ before; so, there is an error of about 0.3 day$^{-1}$ with the autocorrelation method. We show this example to make it clear that the autocorrelation method can only provide an approximation, but it is acceptable for a statistical sample. The modes with similar angular degrees are positioned close to a vertical ridge in the \'echelle diagram; although, the significant variations in the large frequency separation give the diagram the appearance of being bent. With the mode spacing in \dsct stars not being fully regular as the excited modes have radial orders too low to lie in the asymptotic regime, the large separation varies with frequency (especially for lower radial orders \citealp{Garc2009}).

    \begin{figure}
        \centering
        \includegraphics[width=\columnwidth]{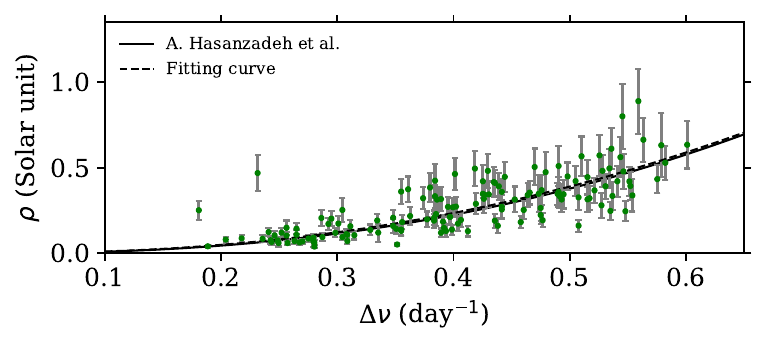}
        \caption{Mean stellar density ($\rho$) versus $\Delta\nu$ (Solar unit) for 148 \dsct stars. The fit curve $\Delta\nu = (0.76 \pm 0.01)\rho^{0.44 \pm 0.01}$ is represented by a black dashed line. The empirical relations $\Delta\nu = (0.76 \pm 0.01)\rho^{0.43 \pm 0.02}$ (black solid line) were obtained by \protect\HAS .}
        \label{figure: nu_rho}
        \end{figure}

    \begin{figure}
        \centering
        \includegraphics[width=\columnwidth]{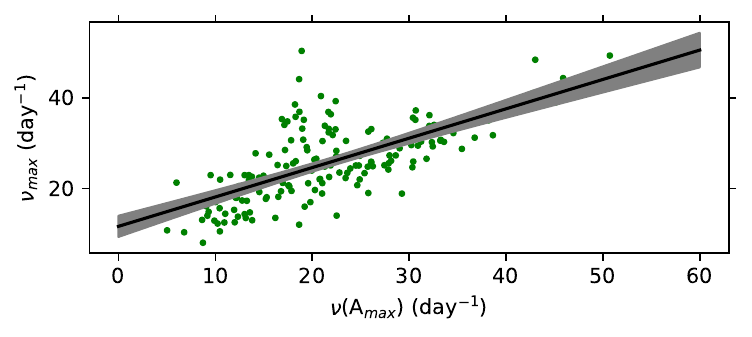}
        \caption{Scatter plot of $\nu_{\mathrm{max}}$ versus $\nu({\rm Amax}$). $\nu_{\mathrm{max}}$ showing a linear relation as $\nu_{\mathrm{max}} = (0.651 \pm 0.033) \nu({\rm Amax}) + (11.622 \pm 0.785)$.}
        \label{figure: nuamax_numax}
        \end{figure}

Figure \ref{figure: fundament} shows the histogram of the fundamental frequency and large frequency separation for 148 \dsct stars. The fundamental frequency and large frequency separation are in the range of $5.01$ to $43.02$ days$^{-1}$ and $2.11$ to $7.33$ days$^{-1}$, respectively. The frequencies are mainly around 10 days$^{-1}$ and 18 days$^{-1}$, with very few stars pulsating at high frequency. The distribution of $\Delta\nu$ is more even between 2.6 days$^{-1}$ and 6.8 days$^{-1}$, except for 4.4 days$^{-1}$. The values of $\Delta\nu$ obtained by \HAS\ are mainly concentrated between 6 days$^{-1}$ and 7 days$^{-1}$. So, there may be no particular pattern to the distribution of $\Delta\nu$ values, which might be a selection effect for different samples.

    \begin{figure}
        \centering
        \includegraphics[width=0.98\columnwidth]{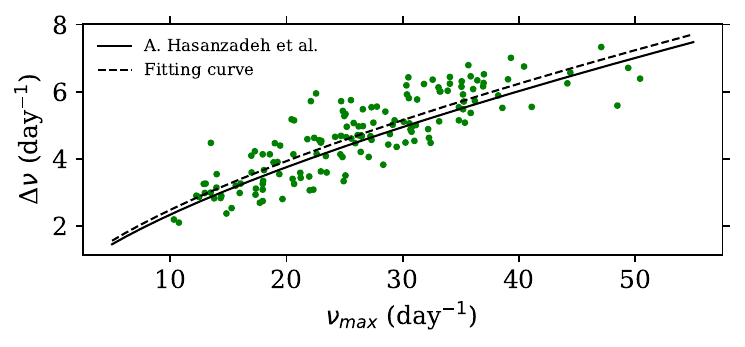}
        \caption{Scatter plot of $\Delta\nu$ versus $\nu_{\mathrm{max}}$. The fit curve $\Delta\nu = (0.53 \pm 0.06)\nu_{\mathrm{max}}^{0.66 \pm 0.03}$ is represented by black dashed line. The empirical relation $\Delta\nu = (0.49 \pm 0.12)\nu_{\mathrm{max}}^{0.68 \pm 0.07}$ (black solid line) was obtained by \protect\HAS .}
        \label{figure: numax_nu}
        \end{figure}

    \begin{figure}
        \centering
        \includegraphics[width=\columnwidth]{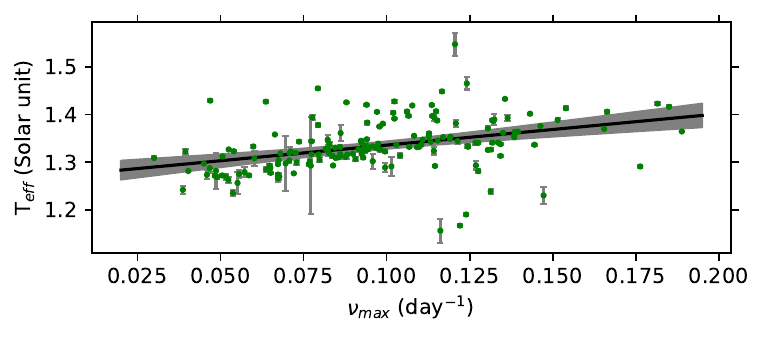}
        \caption{Scatter plot of $\nu_{\mathrm{max}}$ versus T$_{\rm eff}$. T$_{\rm eff}$ shows a linear relation as T$_{\rm eff}$ = $(0.65 \pm 0.08) \nu_{\mathrm{max}} + (1.270 \pm 0.008)$.}
        \label{figure: numax_teff}
        \end{figure}

Figure \ref{figure: nu_rho} depicts the relation between the large separation and stellar density for \dsct stars. A power function $\Delta\nu = (0.76 \pm 0.01)\rho^{0.44 \pm 0.01}$ is fit to data. As we observed in the figure, the fitting is basically consistent with the results of \HAS . According to this fit relation, with the known $\Delta\nu$, we can roughly estimate the density of the star. Determining the $\Delta\nu$ of the star also aids in identifying the frequency modes for a more detailed modeling.

   \begin{figure}
        \centering

        \includegraphics[width=\columnwidth]{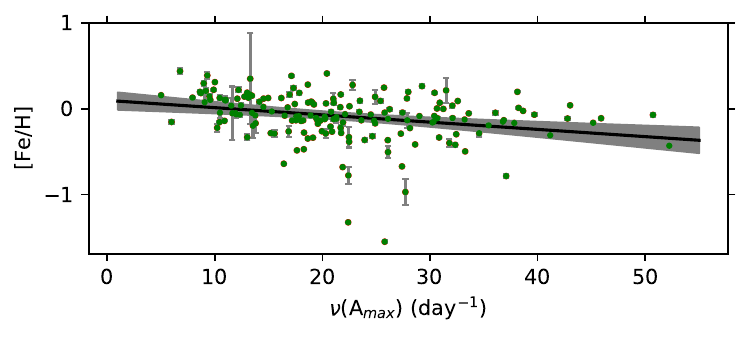}
        \caption{Scatter plot of [Fe/H] versus $\nu({\rm Amax})$. [Fe/H] shows a linear relation as $ \rm [Fe/H]= (-0.008 \pm 0.001) \nu({\rm Amax}) + (0.098 \pm 0.035)$. }
        \label{figure: nuamax_Fe}
        \end{figure}

The scatter plot of $\nu({\rm Amax})$ versus $\nu_{\max}$ is shown in Fig.\ref{figure: nuamax_numax}. $\nu_{\mathrm{max}}$ shows a linear relation, $\nu_{\mathrm{max}} = (0.651 \pm 0.033) \nu({\rm Amax}) + (11.622 \pm 0.785)$. The black lines represent the linear fits of those frequencies using emcee \citep[a Markov chain Monte Carlo Ensemble sampler,][]{Foreman-Mackey2013}. As the temporal evolution of the dominant frequency peak is subject to variability (\HAS ), we chose to employ $\nu_{\max}$ in lieu of the highest amplitude peak for the comparative analysis between $\Delta\nu$ and $\nu_{\max}$, as well as T$_{\rm eff}$ and $\nu_{\max}$. We obtained a power relation, $\Delta\nu = (0.53 \pm 0.06)\nu_{\mathrm{max}}^{0.66 \pm 0.03}$ , for 148 \dsct stars in Fig.\ref{figure: numax_nu}. Also, Fig.\ref{figure: numax_teff} shows the linear correlation T$_{\rm eff}$ = $(0.65 \pm 0.08) \nu_{\mathrm{max}} + (1.270 \pm 0.008)$ obtained through the emcee method for our sample of 179 stars.

According to \citet{Hirschi2008}, stars with lower metallicity are typically hotter, more luminous, and have greater central condensation compared to stars with higher metallicity. This phenomenon is predicted to lead to shorter pulsation periods and higher average densities, as suggested by \citet{Dornan2022}. Fig.\ref{figure: nuamax_Fe} shows a linear relation, $\nu({\rm Amax}) = \rm [Fe/H] - (0.098 \pm 0.035) /  (-0.008 \pm 0.001)$. Based on our results, a decrease in [Fe/H] is related to an increase in pulsation frequencies.

\section{Conclusions}
\label{section: conclusions}

We identified 765 \dsct stars using data from the TESS mission, covering Sectors 40 to 55 with two-minute cadence observations during cycle 4. We compiled two ensembles of \dsct stars parameters using TIC and LAMOST, consisting of 765 and 179 stars, respectively, which we used to investigate \dsct stars pulsational properties. The frequency distribution of diverse parameters, including effective temperature, surface gravity, $V$mag, luminosity, [Fe/H], mass, radius, and parallax, for 765 and 179 \dsct stars. The mass, radius, effective temperature, and surface-gravity values of \dsct stars lie within the ranges of 1.4M$_{\odot}$ to 2.1M$_{\odot}$, 1.2R$_{\odot}$ to 3.7R$_{\odot}$, 6300K to 9300K, and 3.3dex to 4.5dex, respectively, which is consistent with earlier findings on \dsct stars (e.g., \citealp{2007CoAst.150...32D, Antoci2019}). The T$_{\rm eff} - \log g$ diagrams presented in Fig.~\ref{figure: T_logg} reveal that the majority of \dsct stars detected by TESS observations align with the boundaries of the observational instability strip \citep{Murphy2019}, with the subgroup of stars with LAMOST spectral parameters more closely clustered around the theoretical instability strip. Our sample provides confirmation of the observational instability strip derived from Kepler observations and Gaia parameters by \citet{Murphy2019}.

None of the \dsct stars observed by TESS exhibit a maximum pulsation frequency exceeding $60$ days$^{-1}$, primarily due to the absence of hot ZAMS stars, as illustrated in Fig.~\ref{figure: fre_teff}. This result is consistent with earlier studies on \dsct stars using a subset of $Kepler$ data (e.g., \citealp{Balona2011, Bowman2018}). We investigated the relationship between pulsation frequency and T$_{\rm eff}$ by classifying the stars of the TIC sample. After isolating documented Pre-MS \dsct stars, the TIC stars are grouped according to $\log g \geq 4.0$ (ZAMS), $3.5 \leq \log g < 4.0$ (MAMS), and $\log g < 3.5$ (TAMS), as shown in the bottom row of Fig.~\ref{figure: fre_teff}.
According to our statistical results, the T$_{\rm eff}$ and maximum amplitude pulsation frequency of Pre-MS \dsct stars and these three different $\log g$ groups present different fitting results. The fitting results for the ZAMS \dsct stars are in general agreement with the conclusions derived from previous theoretical models and statistical studies. ZAMS \dsct stars have a higher effective temperature and higher radial order of p~modes, leading to frequencies where larger pulsation modes can be observed. The correlation between temperature and the maximum amplitude pulsation frequency is more pronounced among the stars in the MAMS \dsct stars compared to the ZAMS \dsct stars. Due to the limited number of stars in the Pre-MS and TAMS \dsct stars, it is not possible to obtain reliable statistical results. However, based on this trend, it is expected that the TAMS \dsct stars exhibit a more correlated relationship between temperature and the highest amplitude pulsation frequency. Further analysis of Pre-MS and TAMS \dsct stars is necessary to obtain precise and accurate results.

We employed an empirical relation and a 2D autocorrelation method to determine the asteroseismic indices, such as the large frequency separation for each of the 148 \dsct stars and the relationship between these asteroseismic indices. For some of these stars, based on the large frequency separation that has been obtained, it is possible to perform mode identification for the frequencies of these stars. Modeling these most promising stars of our sample, using adapted stellar evolution and oscillation codes that include the effects of rotation, constrains the internal physics of \dsct stars and sheds light on the excitation and selection mechanisms, and on the internal distribution of angular momentum. The large number of photometric data the TESS space telescope provides is a valuable resource that can be used for many years.

We employed an empirical relation and a 2D autocorrelation method to determine the asteroseismic indices, such as the large frequency separation for each of the 148 \dsct stars and the relationship between these asteroseismic indices. For some of these stars, based on the large frequency separation that has been obtained, it is possible to perform mode identification for the frequencies of these stars. Modeling these most promising stars of our sample, using adapted stellar evolution and oscillation codes that include the effects of rotation, constrains the internal physics of \dsct stars and sheds light on the excitation and selection mechanisms, and on the internal distribution of angular momentum. The large number of photometric data the TESS space telescope provides is a valuable resource that can be used for many years.
\begin{acknowledgements}

The authors acknowledge Antonio Garc\'ia Hernandez for a preliminary estimation of asteroseismic indices and gratefully thank the referee for the very useful comments. The work is supported by the National Key R\&D program of China for the Intergovernmental Scientific and Technological Innovation Cooperation Project under No. 2022YFE0126200. Tian- shan Talent Training Program No. 2023TSYCLJ0053. Chenglong Lv would like to acknowledge the funding from the China Scholarship Council scholarship (No 202204910463). A.Hasanzadeh acknowledges support from the Science and Technology Facilities Council (STFC) grant No. ST/X000915/1. JPG acknowledges financial support from project PID2019-107061GB-C63 from the `Programas Estatales de Generaci\'on de Conocimiento y Fortalecimiento Cient\'ifico y Tecnol\'ogico del Sistema de I+D+i y de I+D+i Orientada a los Retos de la Sociedad' and from the Severo Ochoa grant CEX2021-001131-S funded by MCIN/AEI/10.13039/501100011033.
GMM acknowledges funding support from the Spanish State Research Agency (AEI) project PID2019-107061GB-064.
Some of the data presented in this paper were obtained from the Mikulski Archive for Space Telescopes (MAST). The Large Sky Area Multi-Object Fiber Spectroscopic Telescope (LAMOST) is a National Major Scientific Project built by the Chinese Academy of Sciences. Funding for the project has been provided by the National Development and Reform Commission. LAMOST is operated and managed by the National Astronomical Observatories, Chinese Academy of Sciences.

\end{acknowledgements}

\bibliographystyle{aa}
\bibliography{references}

\appendix
\section{Tables}

\begin{landscape}
\begin{table}
\tiny
\centering
\caption{TIC number, positions,$V$mag, effective temperature (T$_{\rm eff}$), Mass ($M/M_{\odot}$), Radius ($R/R_{\odot}$), surface gravity ($\log g$), Luminosity, Parallax(mas),$M_v$, Period (day), Amplitude (mmag), pulsation constants($Q$) for 765 \dsct stars. The complete table is available in machine-readable form at the CDS.}

\begin{tabular}{cccccccccccccc}
\hline
\hline
TIC & RA  & Dec  & $V$mag  & T$_{\rm eff}$   & $M/M_{\odot}$ & $R/R_{\odot}$  & $\log g$   & $L/L_{\odot}$ & Parallax   & $M_v$  & Frequency  & Amplitude & $Q$ \\
 &  (deg) & (deg) & (mag) & (K)   &  &   & (dex)  &  &  (mas)  &  (mag) &  (day$^{-1}$) &  (mmag)&  \\
\hline
1027996	&	154.018774	&	24.73731124	&	11.46 	&	7634 	$\pm$	137 	&	1.78 	$\pm$	0.28 	&	1.31 	$\pm$	0.07 	&	4.46 	$\pm$	0.09 	&	5.21 	$\pm$	0.48 	&	1.73 	$\pm$	0.05 	&	2.65 	&	25.7915 	$\pm$	0.0001 	&	22.91 	$\pm$	0.07 	&	0.031 	\\
1735435	&	197.120911	&	-17.90848965	&	9.68 	&	7301 	$\pm$	138 	&	1.65 	$\pm$	0.28 	&	1.85 	$\pm$	0.08 	&	4.12 	$\pm$	0.09 	&	8.75 	$\pm$	0.56 	&	3.91 	$\pm$	0.03 	&	2.64 	&	18.1629 	$\pm$	0.0003 	&	2.53 	$\pm$	0.04 	&	0.028 	\\
2779096	&	80.652795	&	33.41096325	&	13.44 	&	7547 	$\pm$	72 	&	1.75 		&	2.51 		&	3.88 		&	18.41 		&	0.53 	$\pm$	0.02 	&	2.06 	&	16.4093 	$\pm$	0.0011 	&	2.45 	$\pm$	0.12 	&	0.019 	\\
3373344	&	82.745031	&	32.80356985	&	11.65 	&	6748 	$\pm$	155 	&	1.44 	$\pm$	0.25 	&	3.83 	$\pm$	0.21 	&	3.43 	$\pm$	0.09 	&	27.33 	$\pm$	2.81 	&	1.11 	$\pm$	0.02 	&	1.88 	&	10.7634 	$\pm$	0.0001 	&	21.90 	$\pm$	0.10 	&	0.015 	\\
3571398	&	83.000548	&	34.31183172	&	15.03 	&	6917 	$\pm$	181 	&	1.51 		&	3.22 		&	3.60 		&	21.35 		&	0.44 	$\pm$	0.03 	&	3.27 	&	8.1572 	$\pm$	0.0005 	&	12.39 	$\pm$	0.28 	&	0.018 	\\
3697787	&	127.73742	&	29.78658533	&	10.69 	&	7393 	$\pm$	107 	&	1.69 	$\pm$	0.29 	&	1.74 	$\pm$	0.11 	&	4.18 	$\pm$	0.09 	&	8.17 	$\pm$	0.91 	&	2.11 	$\pm$	0.03 	&	2.31 	&	30.1787 	$\pm$	0.0002 	&	3.38 	$\pm$	0.03 	&	0.017 	\\
3699722	&	127.877165	&	24.0811089	&	5.70 	&	7115 	$\pm$	119 	&	1.58 	$\pm$	0.26 	&	3.08 	$\pm$	0.12 	&	3.66 	$\pm$	0.08 	&	21.93 	$\pm$	0.89 	&	13.62 	$\pm$	0.04 	&	1.37 	&	14.6909 	$\pm$	0.0004 	&	4.76 	$\pm$	0.09 	&	0.015 	\\
3802457	&	169.248944	&	22.48478067	&	10.20 	&	7293 	$\pm$	197 	&	1.65 	$\pm$	0.28 	&	1.99 	$\pm$	0.13 	&	4.06 	$\pm$	0.10 	&	10.08 	$\pm$	0.75 	&	2.39 	$\pm$	0.06 	&	2.09 	&	29.3000 	$\pm$	0.0004 	&	1.42 	$\pm$	0.02 	&	0.015 	\\
4425184	&	346.45769	&	-10.8895374	&	8.07 	&	7539 	$\pm$	131 	&	1.74 	$\pm$	0.27 	&	1.62 	$\pm$	0.06 	&	4.26 	$\pm$	0.08 	&	7.60 	$\pm$	0.33 	&	8.10 	$\pm$	0.03 	&	2.61 	&	18.5645 	$\pm$	0.0003 	&	1.35 	$\pm$	0.02 	&	0.034 	\\
6894037	&	125.550275	&	13.99907263	&	9.77 	&	7298 	$\pm$	14 	&		&		&		&		&	2.49 	$\pm$	0.03 	&	1.75 	&	24.3816 	$\pm$	0.0003 	&	2.75 	$\pm$	0.03 	&		\\
8072655	&	153.858541	&	41.87508017	&	8.24 	&	7057 	$\pm$	116 	&	1.56 	$\pm$	0.27 	&	2.29 	$\pm$	0.09 	&	3.91 	$\pm$	0.08 	&	11.68 	$\pm$	0.53 	&	5.97 	$\pm$	0.04 	&	2.12 	&	6.8536 	$\pm$	0.0004 	&	2.47 	$\pm$	0.05 	&	0.027 	\\
8498156	&	246.719178	&	31.87242841	&	10.58 	&	7339 	$\pm$	203 	&	1.66 	$\pm$	0.27 	&	1.87 	$\pm$	0.11 	&	4.11 	$\pm$	0.10 	&	9.15 	$\pm$	0.41 	&	2.20 	$\pm$	0.01 	&	2.30 	&	16.7751 	$\pm$	0.0004 	&	3.29 	$\pm$	0.05 	&	0.029 	\\
8673205	&	248.756333	&	33.40337656	&	9.78 	&	7976 	$\pm$	135 	&	1.92 	$\pm$	0.29 	&	1.67 	$\pm$	0.05 	&	4.28 	$\pm$	0.07 	&	10.18 	$\pm$	0.41 	&	3.06 	$\pm$	0.01 	&	2.21 	&	16.8718 	$\pm$	0.0003 	&	3.60 	$\pm$	0.05 	&	0.037 	\\
9709684	&	358.205308	&	-6.531686977	&	9.43 	&	7790 	$\pm$	135 	&	1.85 	$\pm$	0.29 	&	1.54 	$\pm$	0.05 	&	4.33 	$\pm$	0.08 	&	7.87 	$\pm$	0.30 	&	4.06 	$\pm$	0.03 	&	2.47 	&	44.3540 	$\pm$	0.0003 	&	1.63 	$\pm$	0.02 	&	0.015 	\\
9958533	&	110.956607	&	38.03115784	&	9.07 	&	6730 	$\pm$	116 	&	1.43 	$\pm$	0.25 	&	2.54 	$\pm$	0.10 	&	3.78 	$\pm$	0.09 	&	11.94 	$\pm$	0.52 	&	4.11 	$\pm$	0.02 	&	2.14 	&	11.9374 	$\pm$	0.0001 	&	3.57 	$\pm$	0.02 	&	0.025 	\\
9960572	&	111.149285	&	37.81141349	&	9.41 	&	7406 	$\pm$	125 	&	1.69 	$\pm$	0.27 	&	1.76 	$\pm$	0.07 	&	4.17 	$\pm$	0.08 	&	8.43 	$\pm$	0.40 	&	4.01 	$\pm$	0.03 	&	2.43 	&	22.1364 	$\pm$	0.0002 	&	2.13 	$\pm$	0.02 	&	0.024 	\\
10289028	&	267.360697	&	34.60574851	&	10.38 	&	7581 	$\pm$	128 	&	1.76 	$\pm$	0.31 	&	1.52 	$\pm$	0.05 	&	4.32 	$\pm$	0.08 	&	6.90 	$\pm$	0.27 	&	2.89 	$\pm$	0.01 	&	2.68 	&	22.1358 	$\pm$	0.0004 	&	1.84 	$\pm$	0.03 	&	0.031 	\\
10359779	&	267.588344	&	34.56240823	&	10.28 	&	7283 	$\pm$	132 	&	1.64 	$\pm$	0.27 	&	2.30 	$\pm$	0.09 	&	3.93 	$\pm$	0.08 	&	13.43 	$\pm$	0.58 	&	2.11 	$\pm$	0.01 	&	1.90 	&	23.4086 	$\pm$	0.0004 	&	2.73 	$\pm$	0.05 	&	0.014 	\\
10490229	&	268.36528	&	32.36800873	&	8.20 	&	7403 	$\pm$	124 	&	1.69 	$\pm$	0.29 	&	1.82 	$\pm$	0.07 	&	4.15 	$\pm$	0.08 	&	8.97 	$\pm$	0.32 	&	6.92 	$\pm$	0.02 	&	2.40 	&	32.9660 	$\pm$	0.0005 	&	1.37 	$\pm$	0.03 	&	0.016 	\\
12594730	&	298.459226	&	27.69167364	&	9.91 	&	9257 	$\pm$	994 	&	2.35 	$\pm$	0.43 	&	2.93 	$\pm$	0.11 	&	3.87 	$\pm$	0.10 	&	56.94 	$\pm$	18.90 	&	1.79 	$\pm$	0.02 	&	1.17 	&	22.4313 	$\pm$	0.0001 	&	7.78 	$\pm$	0.04 	&	0.015 	\\
12653312	&	335.784665	&	-14.21147331	&	8.31 	&	7761 	$\pm$	127 	&	1.83 	$\pm$	0.28 	&	1.85 	$\pm$	0.06 	&	4.17 	$\pm$	0.07 	&	11.13 	$\pm$	0.42 	&	5.98 	$\pm$	0.03 	&	2.19 	&	15.5348 	$\pm$	0.0007 	&	1.55 	$\pm$	0.05 	&	0.034 	\\
12769174	&	336.830163	&	-14.15350313	&	9.52 	&	7564 	$\pm$	131 	&	1.75 	$\pm$	0.29 	&	1.64 	$\pm$	0.06 	&	4.25 	$\pm$	0.08 	&	7.95 	$\pm$	0.39 	&	3.76 	$\pm$	0.02 	&	2.40 	&	29.5318 	$\pm$	0.0007 	&	5.54 	$\pm$	0.03 	&	0.020 	\\
12925840	&	305.074709	&	39.37477779	&	8.89 	&	7751 			&			&	1.96 		&		&		&	5.77 	$\pm$	0.10 	&	2.69 	&	27.6645 	$\pm$	0.0004 	&	0.80 	$\pm$	0.02 	&		\\
13003355	&	305.206727	&	41.21913301	&	9.48 	&	8124 	$\pm$	123 	&	1.98 		&	1.64 		&	4.30 		&	10.58 		&	3.90 	$\pm$	0.01 	&	2.44 	&	40.8996 	$\pm$	0.0004 	&	1.05 	$\pm$	0.02 	&	0.016 	\\
13877064	&	305.745584	&	40.76092111	&	10.64 	&	9279 	$\pm$	124 	&			&	11.66 		&		&		&	0.60 	$\pm$	0.01 	&	-0.47 	&	6.3879 	$\pm$	0.0002 	&	8.68 	$\pm$	0.08 	&		\\
14254276	&	60.16939	&	20.41333384	&	8.04 	&	8207 	$\pm$	424 	&	2.01 	$\pm$	0.34 	&	1.49 	$\pm$	0.05 	&	4.40 	$\pm$	0.08 	&	9.04 	$\pm$	1.70 	&	8.31 	$\pm$	0.03 	&	2.64 	&	21.5390 	$\pm$	0.0002 	&	1.94 	$\pm$	0.02 	&	0.036 	\\
14440714	&	110.687776	&	16.39180665	&	9.99 	&	7628 	$\pm$	138 	&	1.78 	$\pm$	0.28 	&	1.51 	$\pm$	0.05 	&	4.33 	$\pm$	0.08 	&	6.98 	$\pm$	0.28 	&	3.37 	$\pm$	0.02 	&	2.63 	&	18.8707 	$\pm$	0.0001 	&	8.32 	$\pm$	0.05 	&	0.037 	\\
14503989	&	125.459747	&	19.48907444	&	8.76 	&	7146 	$\pm$	184 	&	1.59 	$\pm$	0.27 	&	1.89 	$\pm$	0.10 	&	4.09 	$\pm$	0.09 	&	8.39 	$\pm$	0.34 	&	5.34 	$\pm$	0.02 	&	2.39 	&	26.2025 	$\pm$	0.0002 	&	1.53 	$\pm$	0.01 	&	0.030 	\\
14562583	&	111.189006	&	12.9729295	&	11.63 	&	7007 	$\pm$	144 	&	1.54 	$\pm$	0.28 	&	3.92 	$\pm$	0.29 	&	3.44 	$\pm$	0.11 	&	33.30 	$\pm$	4.40 	&	0.80 	$\pm$	0.04 	&	1.15 	&	7.3108 	$\pm$	0.0002 	&	20.91 	$\pm$	0.14 	&	0.022 	\\
14605806	&	111.280837	&	15.02349858	&	11.92 	&	7679 	$\pm$	131 	&	1.80 	$\pm$	0.29 	&	1.59 	$\pm$	0.08 	&	4.29 	$\pm$	0.09 	&	7.92 	$\pm$	0.69 	&	1.40 	$\pm$	0.02 	&	2.65 	&	37.7694 	$\pm$	0.0005 	&	1.78 	$\pm$	0.04 	&	0.018 	\\
14676022	&	126.859178	&	23.14547614	&	9.79 	&	7794 	$\pm$	128 	&	1.85 	$\pm$	0.29 	&	1.71 	$\pm$	0.06 	&	4.24 	$\pm$	0.08 	&	9.67 	$\pm$	0.40 	&	3.58 	$\pm$	0.02 	&	2.56 	&	37.9472 	$\pm$	0.0023 	&	0.76 	$\pm$	0.08 	&	0.017 	\\
14713971	&	111.940135	&	14.95433012	&	11.52 	&	8147 	$\pm$	123 	&	1.99 	$\pm$	0.30 	&	1.74 	$\pm$	0.07 	&	4.25 	$\pm$	0.08 	&	12.07 	$\pm$	0.82 	&	1.45 	$\pm$	0.02 	&	2.33 	&	33.1333 	$\pm$	0.0006 	&	1.31 	$\pm$	0.04 	&	0.019 	\\
14948284	&	190.282342	&	30.43712189	&	6.95 	&	7178 	$\pm$	130 	&	1.61 	$\pm$	0.28 	&	2.47 	$\pm$	0.10 	&	3.86 	$\pm$	0.09 	&	14.58 	$\pm$	0.58 	&	9.66 	$\pm$	0.02 	&	1.87 	&	10.7532 	$\pm$	0.0003 	&	9.81 	$\pm$	0.16 	&	0.030 	\\
16328463	&	115.972879	&	15.25767509	&	7.27 	&	7160 	$\pm$	150 	&	1.60 	$\pm$	0.28 	&	4.98 	$\pm$	0.23 	&	3.25 	$\pm$	0.09 	&	58.76 	$\pm$	2.62 	&	4.32 	$\pm$	0.03 	&	0.45 	&	6.1913 	$\pm$	0.0002 	&	3.01 	$\pm$	0.03 	&	0.018 	\\
16650539	&	106.366084	&	37.01459538	&	12.01 	&	7610 	$\pm$	135 	&	1.77 	$\pm$	0.29 	&	1.41 	$\pm$	0.06 	&	4.39 	$\pm$	0.08 	&	5.99 	$\pm$	0.33 	&	1.57 	$\pm$	0.02 	&	2.99 	&	22.4137 	$\pm$	0.0003 	&	3.65 	$\pm$	0.05 	&	0.036 	\\
16698135	&	106.546118	&	37.33519725	&	9.29 	&	7222 	$\pm$	129 	&	1.62 	$\pm$	0.27 	&	2.42 	$\pm$	0.10 	&	3.88 	$\pm$	0.09 	&	14.39 	$\pm$	0.71 	&	3.40 	$\pm$	0.02 	&	1.95 	&	13.1291 	$\pm$	0.0004 	&	4.70 	$\pm$	0.10 	&	0.025 	\\
16702228	&	106.623341	&	39.19285262	&	11.23 	&	7614 	$\pm$	147 	&	1.77 	$\pm$	0.29 	&	1.46 	$\pm$	0.06 	&	4.36 	$\pm$	0.08 	&	6.43 	$\pm$	0.42 	&	2.07 	$\pm$	0.02 	&	2.80 	&	22.4012 	$\pm$	0.0007 	&	1.35 	$\pm$	0.04 	&	0.021 	\\
16739247	&	307.950146	&	40.23723199	&	10.23 	&	7706 	$\pm$	179 	&	1.81 	$\pm$	0.30 	&	1.53 	$\pm$	0.05 	&	4.32 	$\pm$	0.08 	&	7.48 	$\pm$	0.52 	&	2.86 	$\pm$	0.02 	&	2.51 	&	22.9062 	$\pm$	0.0003 	&	1.56 	$\pm$	0.02 	&	0.029 	\\
16984366	&	234.677101	&	31.55185002	&	7.22 	&	7537 	$\pm$	125 	&	1.74 	$\pm$	0.27 	&	1.94 	$\pm$	0.06 	&	4.11 	$\pm$	0.08 	&	10.90 	$\pm$	0.32 	&	9.72 	$\pm$	0.02 	&	2.16 	&	26.9527 	$\pm$	0.0003 	&	1.26 	$\pm$	0.02 	&	0.036 	\\
17307872	&	65.536383	&	21.05662191	&	10.88 	&	7587 	$\pm$	101 	&	1.76 	$\pm$	0.27 	&	1.49 	$\pm$	0.05 	&	4.34 	$\pm$	0.08 	&	6.65 	$\pm$	0.27 	&	3.08 	$\pm$	0.02 	&	3.33 	&	30.3946 	$\pm$	0.0003 	&	1.76 	$\pm$	0.02 	&	0.023 	\\
17466526	&	242.619003	&	31.19319058	&	10.42 	&	7610 	$\pm$	98 	&	1.77 	$\pm$	0.28 	&	1.81 	$\pm$	0.05 	&	4.17 	$\pm$	0.08 	&	9.85 	$\pm$	0.39 	&	2.35 	$\pm$	0.02 	&	2.27 	&	15.8801 	$\pm$	0.0021 	&	1.37 	$\pm$	0.10 	&	0.031 	\\
17489989	&	243.127785	&	35.06115312	&	8.88 	&	7449 	$\pm$	131 	&	1.71 	$\pm$	0.28 	&	1.60 	$\pm$	0.06 	&	4.26 	$\pm$	0.08 	&	7.06 	$\pm$	0.24 	&	5.75 	$\pm$	0.02 	&	2.68 	&	33.3832 	$\pm$	0.0002 	&	1.76 	$\pm$	0.02 	&	0.019 	\\
17563488	&	243.866394	&	34.46323684	&	9.48 	&	6916 	$\pm$	131 	&	1.51 	$\pm$	0.26 	&	2.68 	$\pm$	0.11 	&	3.76 	$\pm$	0.09 	&	14.79 	$\pm$	0.71 	&	2.94 	$\pm$	0.03 	&	1.83 	&	10.2700 	$\pm$	0.0002 	&	17.99 	$\pm$	0.13 	&	0.026 	\\
17619425	&	244.668717	&	32.05617824	&	10.80 	&	8086 	$\pm$	144 	&	1.96 	$\pm$	0.29 	&	1.60 	$\pm$	0.07 	&	4.32 	$\pm$	0.08 	&	9.82 	$\pm$	0.56 	&	1.58 	$\pm$	0.05 	&	1.80 	&	30.8549 	$\pm$	0.0002 	&	8.11 	$\pm$	0.06 	&	0.020 	\\
17745208	&	117.8222	&	14.73996528	&	13.93 	&	7259 	$\pm$	140 	&				&				&				&			&	1.32 	$\pm$	0.02 	&	4.54 	&	4.6199 	$\pm$	0.0014 	&	25.29 	$\pm$	1.57 	&		\\
17811089	&	109.628934	&	38.15567022	&	9.69 	&	7336 	$\pm$	132 	&	1.66 	$\pm$	0.26 	&	1.71 	$\pm$	0.06 	&	4.19 	$\pm$	0.08 	&	7.62 	$\pm$	0.33 	&	3.89 	$\pm$	0.02 	&	2.64 	&	20.6128 	$\pm$	0.0002 	&	2.04 	$\pm$	0.02 	&	0.035 	\\
17816363	&	118.392519	&	14.41469636	&	11.60 	&	7586 	$\pm$	131 	&	1.76 	$\pm$	0.30 	&	1.61 	$\pm$	0.07 	&	4.27 	$\pm$	0.09 	&	7.70 	$\pm$	0.61 	&	1.78 	$\pm$	0.05 	&	2.86 	&	21.6974 	$\pm$	0.0004 	&	2.85 	$\pm$	0.05 	&	0.031 	\\
19246219	&	81.823193	&	19.99597278	&	9.45 	&	7124 	$\pm$	135 	&	1.59 	$\pm$	0.28 	&	2.36 	$\pm$	0.36 	&	3.89 	$\pm$	0.16 	&	12.92 	$\pm$	3.72 	&	3.83 	$\pm$	0.03 	&	2.36 	&	14.5352 	$\pm$	0.0004 	&	1.18 	$\pm$	0.02 	&	0.024 	\\
20161974	&	145.300379	&	46.97947628	&	11.25 	&	7413 	$\pm$	110 	&	1.69 	$\pm$	0.26 	&	1.70 	$\pm$	0.08 	&	4.20 	$\pm$	0.08 	&	7.87 	$\pm$	0.61 	&	2.41 	$\pm$	0.05 	&	3.16 	&	20.7627 	$\pm$	0.0004 	&	1.42 	$\pm$	0.03 	&	0.032 	\\

\hline
\end{tabular}
\label{tab:Catalog765}
\end{table}

\end{landscape}

\onecolumn

\tiny
\setlength{\tabcolsep}{12pt}
\begin{longtable}{lccccccccc}

\caption{Catalogs of 179 \dsct stars with LAMOST parameters \citep{Cui2012} and asteroseismic indices.}\label{tab:Catalog179}\\
\hline
\hline
TIC & Spectral type & T$_{\rm eff}$ & Fe/H  &$\log g$ & $Q$     &  $\nu_{f}$      & $\nu(A_{\max})$ & $\nu_{\max}$ & $\Delta\nu$  \\
    &               &(K)           & (dex)    & (dex)     &         & (day$^{-1}$)    &(day$^{-1}$)     &(day$^{-1}$)  & (day$^{-1}$)  \\
\hline
\endfirsthead

\multicolumn{9}{r}{Continue} \\
\hline
\hline
TIC & Spectral type & T$_{\rm eff}$ & Fe/H &$\log g$ & $Q$     &  $\nu_{f}$      & $\nu(A_{\max})$ & $\nu_{\max}$ & $\Delta\nu$  \\
    &               &(K)           & (dex)    & (dex)     &       & (day$^{-1}$)    &(day$^{-1}$)     &(day$^{-1}$)  & (day$^{-1}$)  \\

\hline
\endhead

\hline
\multicolumn{9}{r}{Next page} \\
\endfoot

\hline
\endlastfoot

1027996	    &	A5	&	6509 	$\pm$	20 	&	-1.55 	$\pm$	0.01 	&	4.31 	$\pm$	0.03 	&	0.023 	&	25.00 	&	25.79 	&	32.58 	&	6.36 	\\
2779096	    &	A7	&	7557 	$\pm$	86 	&	-0.15 	$\pm$	0.08 	&	3.9 	$\pm$	0.1 	&	0.020 	&		&		&		&		\\
3697787	    &	F0	&	7398 	$\pm$	23 	&	0.06 	$\pm$	0.01 	&	3.94 	$\pm$	0.03 	&	0.013 	&	15.79 	&	20.34 	&	26.21 	&	4.97 	\\
6894037	    &	F0	&	7274 	$\pm$	16 	&	0.05 	$\pm$	0.01 	&	3.94 	$\pm$	0.02 	&	0.012 	&		&		&		&		\\
8498156	    &	F0	&	7358 	$\pm$	15 	&	0.02 	$\pm$	0.01 	&	3.96 	$\pm$	0.02 	&	0.024 	&	16.47 	&	16.78 	&	19.39 	&	3.55 	\\
8673205	    &	A7	&	7748 	$\pm$	67 	&	-0.26 	$\pm$	0.07 	&	4.21 	$\pm$	0.09 	&	0.033 	&	14.72 	&	16.87 	&	35.35 	&	5.08 	\\
9958533	    &	F0	&	7132 	$\pm$	59 	&	-0.05 	$\pm$	0.06 	&	3.96 	$\pm$	0.08 	&	0.032 	&	10.93 	&	11.94 	&	15.28 	&	2.54 	\\
16650539	&	A3	&	6862 	$\pm$	96 	&	-0.8 	$\pm$	0.1 	&	4.2 	$\pm$	0.2 	&	0.027 	&	22.36 	&	22.42 	&	39.31 	&	7.01 	\\
16702228	&	A6	&	6639 	$\pm$	13 	&	-1.32 	$\pm$	0.01 	&	4.34 	$\pm$	0.02 	&	0.018 	&	19.72 	&	22.40 	&	33.08 	&	6.13 	\\
17307872	&	A7	&	7480 	$\pm$	10 	&	-0.09 	$\pm$	0.01 	&	3.97 	$\pm$	0.01 	&	0.015 	&	30.40 	&	30.40 	&	35.65 	&	6.79 	\\
17489989	&	A7	&	7382 	$\pm$	42 	&	0.07 	$\pm$	0.02 	&	4.01 	$\pm$	0.05 	&	0.014 	&	18.83 	&	21.06 	&	30.49 	&	6.43 	\\
17619425	&	A6	&	8078 	$\pm$	17 	&	-0.34 	$\pm$	0.01 	&	4.25 	$\pm$	0.02 	&	0.018 	&	28.26 	&	30.85 	&	31.14 	&	5.01 	\\
17811089	&	F0	&	7317 	$\pm$	19 	&	0.13 	$\pm$	0.01 	&	4.00 	$\pm$	0.02 	&	0.028 	&	7.27 	&	16.19 	&	13.50 	&	4.48 	\\
17816363	&	A7	&	7496 	$\pm$	35 	&	-0.22 	$\pm$	0.03 	&	4.17 	$\pm$	0.05 	&	0.027 	&	21.46 	&	21.70 	&	32.40 	&	4.48 	\\
19246219	&	F0	&	7119 	$\pm$	9 	&	0.12 	$\pm$	0.01 	&	3.97 	$\pm$	0.01 	&	0.026 	&	14.54 	&	14.54 	&	19.23 	&	3.91 	\\
20161974	&	A7	&	7358 	$\pm$	14 	&	-0.21 	$\pm$	0.01 	&	4.11 	$\pm$	0.02 	&	0.028 	&	20.71 	&	20.76 	&	22.05 	&		\\
21176331	&	A7	&	7348 	$\pm$	13 	&	-0.26 	$\pm$	0.01 	&	4.14 	$\pm$	0.02 	&	0.029 	&	19.26 	&	19.96 	&	23.98 	&		\\
21631554	&	A7	&	7454 	$\pm$	11 	&	-0.14 	$\pm$	0.01 	&	4.02 	$\pm$	0.01 	&	0.025 	&	17.40 	&	18.22 	&	38.58 	&	5.52 	\\
22685627	&	F0	&	7106 	$\pm$	14 	&	0.18 	$\pm$	0.01 	&	4.15 	$\pm$	0.02 	&	0.049 	&		&		&		&		\\
26694621	&	A6	&	7934 	$\pm$	20 	&	0.04 	$\pm$	0.01 	&	3.90 	$\pm$	0.03 	&	0.009 	&	43.02 	&	43.02 	&	48.47 	&	5.59 	\\
27420275	&	F0	&	7399 	$\pm$	8 	&	0.03 	$\pm$	0.01 	&	3.89 	$\pm$	0.01 	&	0.014 	&	8.55 	&	22.52 	&	14.00 	&	3.55 	\\
43789560	&	A6	&	7326 	$\pm$	25 	&	0.08 	$\pm$	0.02 	&	3.90 	$\pm$	0.04 	&	0.025 	&	12.00 	&	14.15 	&	27.78 	&	5.56 	\\
46479551	&	F0	&	7233 	$\pm$	17 	&	0.28 	$\pm$	0.01 	&	3.88 	$\pm$	0.02 	&	0.020 	&	8.33 	&	18.65 	&	12.00 	&		\\
54766928	&	A3	&	7826 	$\pm$	18 	&	-0.17 	$\pm$	0.02 	&	4.09 	$\pm$	0.03 	&	0.021 	&	7.93 	&	24.91 	&	27.23 	&	4.68 	\\
57333193	&	A7	&	7261 	$\pm$	89 	&	-0.13 	$\pm$	0.09 	&	4.0 	$\pm$	0.1 	&	0.015 	&	21.73 	&	27.89 	&	25.58 	&	4.61 	\\
58994538	&	A6	&	7957 	$\pm$	14 	&	-0.14 	$\pm$	0.01 	&	3.91 	$\pm$	0.02 	&	0.015 	&	15.72 	&	19.81 	&	16.98 	&	4.10 	\\
59081460	&	A7	&	7411 	$\pm$	5 	&	-0.10 	$\pm$	0.01 	&	4.07 	$\pm$	0.01 	&	0.019 	&	21.22 	&	21.30 	&	24.85 	&	5.43 	\\
59938994	&	A7	&	7543 	$\pm$	14 	&	-0.15 	$\pm$	0.01 	&	4.02 	$\pm$	0.02 	&	0.013 	&	19.01 	&	36.77 	&	31.24 	&	5.77 	\\
59944386	&	A7	&	7399 	$\pm$	20 	&	-0.64 	$\pm$	0.01 	&	4.29 	$\pm$	0.03 	&	0.037 	&	16.35 	&	16.42 	&	25.19 	&	4.96 	\\
62978056	&	F0	&	7326 	$\pm$	24 	&	-0.04 	$\pm$	0.02 	&	4.08 	$\pm$	0.03 	&	0.019 	&	12.42 	&	23.44 	&	22.32 	&	3.09 	\\
63003749	&	A7	&	7523 	$\pm$	11 	&	-0.03 	$\pm$	0.01 	&	4.01 	$\pm$	0.01 	&	0.012 	&	20.31 	&	38.65 	&	31.78 	&		\\
63208192	&	A7	&	7713 	$\pm$	22 	&	-0.05 	$\pm$	0.02 	&	3.87 	$\pm$	0.04 	&	0.012 	&	19.14 	&	27.45 	&	25.14 	&	4.65 	\\
66670807	&	F0	&	7286 	$\pm$	21 	&	0.12 	$\pm$	0.01 	&	4.06 	$\pm$	0.03 	&	0.026 	&	8.81 	&	27.86 	&	24.19 	&	4.66 	\\
68541960	&	A7	&	7593 	$\pm$	7 	&	-0.17 	$\pm$	0.01 	&	4.03 	$\pm$	0.01 	&	0.012 	&	20.55 	&	37.83 	&	36.06 	&	6.08 	\\
69030207	&	A7	&	7503 	$\pm$	10 	&	0.07 	$\pm$	0.01 	&	3.96 	$\pm$	0.01 	&	0.021 	&	16.50 	&	19.07 	&	30.80 	&	4.96 	\\
69299060	&	A5	&	8629 	$\pm$	133 &	0.2 	$\pm$	0.2 	&	3.6 	$\pm$	0.2 	&	0.009 	&	31.43 	&	31.50 	&	32.19 	&	4.89 	\\
69675850	&	A7	&	7645 	$\pm$	24 	&	-0.05 	$\pm$	0.02 	&	3.96 	$\pm$	0.04 	&	0.012 	&	19.11 	&	36.08 	&	34.83 	&	5.15 	\\
70547124	&	A7	&	7324 	$\pm$	14 	&	-0.47 	$\pm$	0.01 	&	4.15 	$\pm$	0.02 	&	0.030 	&	18.21 	&	18.28 	&	35.85 	&	6.46 	\\
79408364	&	A7	&	7227 	$\pm$	13 	&	0.12 	$\pm$	0.01 	&	3.90 	$\pm$	0.02 	&	0.026 	&	12.10 	&	12.10 	&	17.96 	&	2.75 	\\
80293642	&	F0	&	7114 	$\pm$	45 	&	0.39 	$\pm$	0.04 	&	3.79 	$\pm$	0.06 	&	0.031 	&	9.25 	&	9.31 	&	14.84 	&	2.38 	\\
82202857	&	A7	&	7477 	$\pm$	31 	&	0.04 	$\pm$	0.02 	&	3.99 	$\pm$	0.04 	&	0.014 	&	17.09 	&	32.08 	&	33.83 	&		\\
82351638	&	A6	&	7880 	$\pm$	23 	&	-0.11 	$\pm$	0.02 	&	4.05 	$\pm$	0.04 	&	0.011 	&	25.53 	&	42.78 	&	41.10 	&	5.55 	\\
82452397	&	F0	&	7404 	$\pm$	21 	&	-0.01 	$\pm$	0.02 	&	4.07 	$\pm$	0.03 	&	0.015 	&	20.83 	&	26.25 	&	24.92 	&	3.34 	\\
83283121	&	A7	&	7490 	$\pm$	20 	&	-0.11 	$\pm$	0.02 	&	3.92 	$\pm$	0.03 	&	0.017 	&	20.00 	&	20.28 	&	19.66 	&	2.81 	\\
84132863	&	A7	&	7969 	$\pm$	14 	&	-0.07 	$\pm$	0.01 	&	3.80 	$\pm$	0.02 	&	0.020 	&	12.00 	&	12.00 	&	12.52 	&		\\
84598340	&	A7	&	7766 	$\pm$	39 	&	-0.16 	$\pm$	0.04 	&	4.00 	$\pm$	0.06 	&	0.021 	&	17.14 	&	21.92 	&	36.42 	&	6.34 	\\
84910808	&	A5	&	7950 	$\pm$	8 	&	-0.13 	$\pm$	0.01 	&	4.08 	$\pm$	0.01 	&	0.016 	&	16.42 	&	23.62 	&	23.46 	&	3.60 	\\
86269618	&	A7	&	7453 	$\pm$	42 	&	-0.29 	$\pm$	0.04 	&	4.14 	$\pm$	0.06 	&	0.033 	&	5.02 	&	15.56 	&	27.48 	&	5.08 	\\
86730101	&	A6	&	7772 	$\pm$	30 	&	-0.32 	$\pm$	0.03 	&	4.08 	$\pm$	0.04 	&	0.020 	&	6.41 	&	24.65 	&	20.73 	&		\\
86955845	&	A7	&	7514 	$\pm$	10 	&	-0.16 	$\pm$	0.01 	&	4.01 	$\pm$	0.01 	&	0.015 	&	14.24 	&	30.23 	&	29.60 	&		\\
87209791	&	A7	&	7558 	$\pm$	11 	&	-0.09 	$\pm$	0.01 	&	3.96 	$\pm$	0.01 	&	0.013 	&	29.02 	&	29.02 	&	28.90 	&	4.74 	\\
87499033	&	A7	&	7573 	$\pm$	9 	&	-0.28 	$\pm$	0.01 	&	4.07 	$\pm$	0.01 	&	0.030 	&	12.49 	&	15.21 	&	17.71 	&	2.70 	\\
87698769	&	A7	&	7438 	$\pm$	20 	&	0.17 	$\pm$	0.02 	&	3.95 	$\pm$	0.03 	&	0.022 	&	21.42 	&	21.71 	&	33.19 	&	6.01 	\\
90942000	&	A7	&	7638 	$\pm$	13 	&	-0.35 	$\pm$	0.01 	&	4.15 	$\pm$	0.02 	&	0.030 	&	18.22 	&	18.65 	&	44.17 	&	6.25 	\\
91026282	&	F0	&	7341 	$\pm$	34 	&	0.17 	$\pm$	0.03 	&	3.97 	$\pm$	0.05 	&	0.025 	&	15.64 	&	16.96 	&	21.23 	&	3.44 	\\
91678673	&	A7	&	7467 	$\pm$	15 	&	-0.28 	$\pm$	0.01 	&	4.13 	$\pm$	0.02 	&	0.031 	&	18.30 	&	18.30 	&	26.03 	&		\\
91741040	&	A7	&	7422 	$\pm$	9 	&	-0.12 	$\pm$	0.01 	&	4.03 	$\pm$	0.01 	&	0.019 	&	20.77 	&	26.10 	&	25.78 	&	5.07 	\\
95167223	&	F0	&	7310 	$\pm$	25 	&	0.09 	$\pm$	0.02 	&	3.92 	$\pm$	0.04 	&	0.014 	&	11.28 	&	25.40 	&	23.39 	&	4.09 	\\
97363314	&	F0	&	7434 	$\pm$	16 	&	0.05 	$\pm$	0.01 	&	3.89 	$\pm$	0.02 	&	0.017 	&	19.17 	&	19.22 	&	15.98 	&	2.99 	\\
97640857	&	A3	&	8170 	$\pm$	71 	&	-0.51 	$\pm$	0.07 	&	4.18 	$\pm$	0.12 	&	0.021 	&	15.42 	&	26.10 	&	33.14 	&	5.12 	\\
99034835	&	F0	&	7345 	$\pm$	18 	&	0.19 	$\pm$	0.01 	&	3.92 	$\pm$	0.02 	&	0.026 	&	11.44 	&	13.02 	&	22.96 	&	4.49 	\\
99237171	&	A7	&	7667 	$\pm$	8 	&	-0.23 	$\pm$	0.01 	&	4.00 	$\pm$	0.01 	&	0.015 	&	10.85 	&	28.14 	&	26.12 	&	4.71 	\\
99893404	&	A5	&	7188 	$\pm$	48 	&	-0.40 	$\pm$	0.05 	&	4.14 	$\pm$	0.06 	&	0.014 	&	13.43 	&	31.80 	&	26.57 	&	5.48 	\\
101537683	&	A6	&	7842 	$\pm$	12 	&	-0.39 	$\pm$	0.01 	&	4.12 	$\pm$	0.02 	&	0.020 	&	21.04 	&	22.49 	&	28.33 	&	3.83 	\\
101679169	&	A7	&	7500 	$\pm$	10 	&	-0.09 	$\pm$	0.01 	&	4.03 	$\pm$	0.01 	&	0.029 	&	6.81 	&	20.22 	&	24.64 	&	4.14 	\\
114597980	&	A6	&	7990 	$\pm$	12 	&	-0.11 	$\pm$	0.01 	&	3.99 	$\pm$	0.02 	&	0.014 	&	15.89 	&	32.10 	&	36.21 	&	5.73 	\\
115070770	&	F0	&	7684 	$\pm$	26 	&	-0.12 	$\pm$	0.02 	&	3.97 	$\pm$	0.04 	&	0.020 	&	18.15 	&	21.04 	&	21.19 	&	3.59 	\\
115293110	&	F2	&	7061 	$\pm$	39 	&	0.30 	$\pm$	0.03 	&	3.73 	$\pm$	0.05 	&	0.023 	&	8.98 	&	8.98 	&	17.96 	&	3.26 	\\
115510161	&	A7	&	7209 	$\pm$	564 &	0.4 	$\pm$	0.5 	&	3.83 	$\pm$	0.88 	&	0.022 	&	10.81 	&	13.30 	&	20.60 	&	4.14 	\\
118303902	&	F0	&	7300 	$\pm$	11 	&	-0.08 	$\pm$	0.01 	&	3.97 	$\pm$	0.01 	&	0.029 	&	10.66 	&	13.77 	&	22.60 	&	4.15 	\\
126322311	&	A6	&	7838 	$\pm$	15 	&	-0.37 	$\pm$	0.01 	&	4.13 	$\pm$	0.02 	&	0.018 	&	12.51 	&	26.08 	&	25.92 	&	4.47 	\\
127109148	&	A8	&	7185 	$\pm$	14 	&	0.14 	$\pm$	0.01 	&	3.87 	$\pm$	0.02 	&	0.026 	&	11.01 	&	12.77 	&	17.34 	&	2.94 	\\
130200644	&	A7	&	7478 	$\pm$	10 	&	0.00 	$\pm$	0.01 	&	4.01 	$\pm$	0.01 	&	0.013 	&	18.21 	&	30.65 	&	35.19 	&	5.92 	\\
137214324	&	F0	&	7338 	$\pm$	56 	&	0.28 	$\pm$	0.05 	&	3.91 	$\pm$	0.07 	&	0.019 	&	10.31 	&	22.81 	&	23.53 	&		\\
138285753	&	A7	&	7541 	$\pm$	17 	&	0.07 	$\pm$	0.02 	&	3.93 	$\pm$	0.03 	&	0.022 	&	16.97 	&	18.68 	&	36.96 	&	6.26 	\\
138566633	&	A7	&	7918 	$\pm$	14 	&	-0.05 	$\pm$	0.01 	&	3.96 	$\pm$	0.02 	&	0.012 	&	15.66 	&	33.60 	&	30.32 	&	6.19 	\\
139201927	&	A6	&	7235 	$\pm$	321 &	-0.1 	$\pm$	0.3 	&	3.9 	$\pm$	0.5 	&	0.029 	&	10.49 	&	11.64 	&	18.57 	&	4.14 	\\
139381224	&	F0	&	7378 	$\pm$	11 	&	0.11 	$\pm$	0.01 	&	3.85 	$\pm$	0.01 	&	0.027 	&	8.01 	&	11.02 	&	14.41 	&	2.90 	\\
154485903	&	A5	&	7177 	$\pm$	74 	&	-0.77 	$\pm$	0.07 	&	4.2 	$\pm$	0.1 	&	0.029 	&		&		&		&		\\
155451763	&	F0	&	7984 	$\pm$	236 &	0.1 	$\pm$	0.3 	&	4.5 	$\pm$	0.4 	&	0.020 	&		&		&		&		\\
155457396	&	F0	&	7405 	$\pm$	10 	&	-0.05 	$\pm$	0.01 	&	3.93 	$\pm$	0.01 	&	0.027 	&	10.71 	&	13.46 	&	21.80 	&	4.59 	\\
157521786	&	A7	&	7476 	$\pm$	12 	&	0.09 	$\pm$	0.01 	&	3.94 	$\pm$	0.02 	&	0.013 	&	21.21 	&	32.58 	&	34.06 	&	6.44 	\\
157624150	&	A6	&	7356 	$\pm$	68 	&	0.12 	$\pm$	0.06 	&	4.2 	$\pm$	0.1 	&	0.023 	&	17.12 	&	21.02 	&	18.89 	&	3.90 	\\
157674759	&	F0	&	7375 	$\pm$	34 	&	0.13 	$\pm$	0.03 	&	3.97 	$\pm$	0.05 	&	0.041 	&	10.40 	&	10.47 	&	10.52 	&		\\
157840360	&	A7	&	7460 	$\pm$	13 	&	0.20 	$\pm$	0.01 	&	3.97 	$\pm$	0.02 	&	0.011 	&	19.28 	&	38.12 	&	35.85 	&	5.37 	\\
157947269	&	A7	&	7554 	$\pm$	13 	&	-0.29 	$\pm$	0.01 	&	4.04 	$\pm$	0.02 	&	0.015 	&	13.15 	&	27.29 	&	30.08 	&	5.11 	\\
158175593	&	F0	&	7270 	$\pm$	29 	&	0.26 	$\pm$	0.02 	&	3.86 	$\pm$	0.04 	&	0.023 	&	14.19 	&	29.26 	&	18.87 	&		\\
159101753	&	A5	&	7703 	$\pm$	41 	&	-0.29 	$\pm$	0.04 	&	4.06 	$\pm$	0.06 	&	0.013 	&	27.19 	&	34.57 	&	32.26 	&	4.63 	\\
159582964	&	A7	&	7921 	$\pm$	11 	&	-0.07 	$\pm$	0.01 	&	3.89 	$\pm$	0.02 	&	0.014 	&	22.82 	&	24.50 	&	25.08 	&	5.34 	\\
161029401	&	A7	&	7403 	$\pm$	29 	&	-0.36 	$\pm$	0.03 	&	4.16 	$\pm$	0.05 	&	0.023 	&	17.80 	&	23.93 	&	24.42 	&		\\
162576798	&	A6	&	6450 	$\pm$	141 &	-0.9 	$\pm$	0.2 	&	4.1 	$\pm$	0.2 	&	0.013 	&	15.33 	&	27.72 	&	31.01 	&	4.54 	\\
165286840	&	A9	&	7076 	$\pm$	25 	&	0.20 	$\pm$	0.02 	&	3.74 	$\pm$	0.04 	&	0.029 	&	7.33 	&	8.63 	&	13.05 	&	3.27 	\\
165917502	&	F0	&	7179 	$\pm$	50 	&	0.10 	$\pm$	0.05 	&	3.62 	$\pm$	0.07 	&	0.023 	&	9.90 	&	10.93 	&	12.48 	&	2.87 	\\
166656852	&	F0	&	6890 	$\pm$	39 	&	-0.33 	$\pm$	0.03 	&	4.22 	$\pm$	0.06 	&	0.035 	&	5.62 	&	13.02 	&	14.34 	&	2.84 	\\
171732086	&	A8	&	7095 	$\pm$	15 	&	0.13 	$\pm$	0.01 	&	3.88 	$\pm$	0.02 	&	0.024 	&	8.03 	&	13.14 	&	13.48 	&	3.00 	\\
172171384	&	A6	&	7897 	$\pm$	24 	&	-0.07 	$\pm$	0.02 	&	3.99 	$\pm$	0.03 	&	0.008 	&	40.69 	&	50.72 	&	49.40 	&	6.71 	\\
172825030	&	A7	&	7590 	$\pm$	95 	&	0.04 	$\pm$	0.09 	&	3.9 	$\pm$	0.2 	&	0.028 	&	5.30 	&	11.54 	&	23.00 	&	4.54 	\\
176061004	&	A6	&	6908 	$\pm$	27 	&	-0.78 	$\pm$	0.02 	&	4.23 	$\pm$	0.03 	&	0.013 	&	17.03 	&	37.08 	&	35.06 	&	6.15 	\\
184784453	&	A7	&	7420 	$\pm$	14 	&	-0.28 	$\pm$	0.01 	&	4.11 	$\pm$	0.02 	&	0.025 	&	21.68 	&	21.75 	&	22.55 	&	5.95 	\\
184847294	&	A7	&	7410 	$\pm$	15 	&	0.19 	$\pm$	0.01 	&	3.97 	$\pm$	0.02 	&	0.014 	&	12.66 	&	30.43 	&	25.96 	&		\\
188389973	&	A7	&	7397 	$\pm$	12 	&	0.01 	$\pm$	0.01 	&	4.02 	$\pm$	0.02 	&	0.012 	&	30.18 	&	38.23 	&	35.08 	&	6.31 	\\
188633014	&	A7	&	7340 	$\pm$	21 	&	-0.13 	$\pm$	0.02 	&	4.11 	$\pm$	0.03 	&	0.028 	&	17.48 	&	17.55 	&	20.66 	&	5.15 	\\
193261746	&	A7	&	7840 	$\pm$	23 	&	-0.11 	$\pm$	0.02 	&	3.92 	$\pm$	0.03 	&	0.009 	&	33.28 	&	45.90 	&	44.42 	&	6.57 	\\
197179793	&	A7	&	7426 	$\pm$	9 	&	-0.11 	$\pm$	0.01 	&	4.03 	$\pm$	0.01 	&	0.024 	&	19.38 	&	19.43 	&	29.14 	&	5.02 	\\
197184634	&	F0	&	7084 	$\pm$	43 	&	0.22 	$\pm$	0.03 	&	3.87 	$\pm$	0.06 	&	0.027 	&	12.16 	&	12.21 	&	17.96 	&	4.14 	\\
197185879	&	A7	&	7845 	$\pm$	15 	&	-0.09 	$\pm$	0.01 	&	3.93 	$\pm$	0.02 	&	0.023 	&	16.32 	&	17.82 	&	30.67 	&	4.88 	\\
199712197	&	A6	&	7740 	$\pm$	14 	&	-0.34 	$\pm$	0.01 	&	4.15 	$\pm$	0.02 	&	0.027 	&	19.05 	&	19.13 	&	35.19 	&	5.48 	\\
224333256	&	A7	&	7427 	$\pm$	20 	&	0.15 	$\pm$	0.01 	&	3.98 	$\pm$	0.03 	&	0.031 	&	13.31 	&	13.46 	&	22.96 	&	3.63 	\\
232625872	&	A7	&	7771 	$\pm$	38 	&	0.09 	$\pm$	0.03 	&	4.00 	$\pm$	0.05 	&	0.019 	&	23.41 	&	23.49 	&	30.53 	&	5.81 	\\
238578926	&	A7	&	7212 	$\pm$	47 	&	-0.12 	$\pm$	0.04 	&	4.02 	$\pm$	0.07 	&	0.020 	&	19.19 	&	21.30 	&	33.87 	&	6.03 	\\
239281329	&	F0	&	7453 	$\pm$	7 	&	0.07 	$\pm$	0.01 	&	3.93 	$\pm$	0.01 	&	0.013 	&	14.24 	&	30.91 	&	29.48 	&	4.36 	\\
239973107	&	F0	&	7106 	$\pm$	20 	&	0.10 	$\pm$	0.01 	&	3.89 	$\pm$	0.03 	&	0.030 	&	11.01 	&	11.01 	&	17.96 	&	3.09 	\\
247411272	&	A5	&	7737 	$\pm$	13 	&	-0.50 	$\pm$	0.01 	&	4.14 	$\pm$	0.02 	&	0.016 	&	26.26 	&	33.27 	&	30.76 	&	4.81 	\\
252904295	&	F0	&	7083 	$\pm$	38 	&	-0.03 	$\pm$	0.03 	&	4.04 	$\pm$	0.05 	&	0.027 	&	5.25 	&	15.30 	&	18.07 	&	3.67 	\\
252986944	&	A6	&	7295 	$\pm$	84 	&	0.08 	$\pm$	0.08 	&	3.9 	$\pm$	0.1 	&	0.030 	&	9.02 	&	9.07 	&	16.07 	&	3.27 	\\
255733180	&	A6	&	7275 	$\pm$	8 	&	-0.14 	$\pm$	0.01 	&	4.04 	$\pm$	0.01 	&	0.035 	&	8.47 	&	10.91 	&	18.00 	&	3.35 	\\
262589713	&	A6	&	7306 	$\pm$	0 	&	0.13 	$\pm$	0.01 	&	3.93 	$\pm$	0.01 	&	0.025 	&	14.83 	&	14.97 	&	22.82 	&	4.56 	\\
265858004	&	A5	&	7609 	$\pm$	4 	&	-0.08 	$\pm$	0.01 	&	4.07 	$\pm$	0.01 	&	0.027 	&	17.66 	&	18.91 	&	50.42 	&	6.39 	\\
270214133	&	F0	&	7304 	$\pm$	11 	&	0.25 	$\pm$	0.01 	&	3.91 	$\pm$	0.01 	&	0.015 	&	15.33 	&	25.73 	&	24.82 	&	4.06 	\\
277673440	&	F0	&	7080 	$\pm$	35 	&	-0.07 	$\pm$	0.03 	&	4.08 	$\pm$	0.05 	&	0.033 	&	12.03 	&	12.33 	&	13.78 	&	2.83 	\\
284023292	&	F0	&	7431 	$\pm$	10 	&	0.08 	$\pm$	0.01 	&	3.92 	$\pm$	0.01 	&	0.022 	&	17.48 	&	18.97 	&	33.24 	&	6.01 	\\
287135197	&	A7	&	7429 	$\pm$	11 	&	-0.14 	$\pm$	0.01 	&	3.96 	$\pm$	0.01 	&	0.024 	&	17.18 	&	17.19 	&	28.51 	&	5.41 	\\
287254034	&	A7	&	7489 	$\pm$	26 	&	-0.42 	$\pm$	0.02 	&	4.15 	$\pm$	0.04 	&	0.015 	&	23.66 	&	32.35 	&	30.24 	&	4.49 	\\
289527518	&	A7	&	7494 	$\pm$	13 	&	-0.09 	$\pm$	0.01 	&	3.94 	$\pm$	0.02 	&	0.018 	&	10.68 	&	17.58 	&	20.66 	&	3.26 	\\
295698403	&	A7	&	7512 	$\pm$	60 	&	-0.05 	$\pm$	0.06 	&	3.9 	$\pm$	0.1 	&	0.020 	&	10.37 	&	10.49 	&	21.96 	&	3.48 	\\
302264415	&	F0	&	7149 	$\pm$	207 &	-0.2 	$\pm$	0.1 	&	3.92 	$\pm$	0.26 	&	0.026 	&	6.50 	&	13.80 	&	12.98 	&	2.99 	\\
302654026	&	A8	&	7147 	$\pm$	12 	&	0.16 	$\pm$	0.01 	&	3.98 	$\pm$	0.02 	&	0.085 	&	5.01 	&	5.02 	&	10.75 	&	2.11 	\\
302820047	&	F0	&	7323 	$\pm$	33 	&	0.15 	$\pm$	0.03 	&	4.00 	$\pm$	0.05 	&	0.016 	&	6.94 	&	9.51 	&	22.96 	&	4.50 	\\
302906300	&	F0	&	7375 	$\pm$	22 	&	0.41 	$\pm$	0.01 	&	3.82 	$\pm$	0.03 	&	0.018 	&	19.95 	&	20.42 	&	26.58 	&	4.98 	\\
308303198	&	A7	&	7258 	$\pm$	26 	&	0.04 	$\pm$	0.02 	&	3.94 	$\pm$	0.04 	&	0.030 	&	10.49 	&	12.44 	&	20.53 	&	3.41 	\\
309660658	&	F0	&	7167 	$\pm$	26 	&	0.31 	$\pm$	0.02 	&	3.79 	$\pm$	0.03 	&	0.026 	&	9.59 	&	10.02 	&	17.00 	&	3.60 	\\
312691592	&	A7	&	7444 	$\pm$	21 	&	-0.14 	$\pm$	0.02 	&	4.00 	$\pm$	0.03 	&	0.015 	&	21.30 	&	24.81 	&	25.08 	&	3.51 	\\
318967742	&	A7	&	7201 	$\pm$	113 &	-0.3 	$\pm$	0.1 	&	4.1 	$\pm$	0.2 	&	0.019 	&	15.28 	&	22.47 	&	27.08 	&	4.06 	\\
320419098	&	A5	&	7518 	$\pm$	10 	&	-0.68 	$\pm$	0.01 	&	4.22 	$\pm$	0.01 	&	0.031 	&	19.13 	&	21.89 	&	25.10 	&		\\
330604356	&	F0	&	7386 	$\pm$	26 	&	0.24 	$\pm$	0.02 	&	3.89 	$\pm$	0.04 	&	0.021 	&	16.20 	&	17.32 	&	24.98 	&	5.28 	\\
337264872	&	A7	&	7460 	$\pm$	24 	&	-0.04 	$\pm$	0.02 	&	3.98 	$\pm$	0.03 	&	0.020 	&	18.86 	&	20.82 	&	22.10 	&	5.72 	\\
350498608	&	F0	&	7247 	$\pm$	29 	&	0.19 	$\pm$	0.02 	&	3.83 	$\pm$	0.04 	&	0.020 	&	15.61 	&	17.86 	&	19.46 	&	4.40 	\\
353927180	&	F0	&	7126 	$\pm$	11 	&	0.10 	$\pm$	0.01 	&	3.83 	$\pm$	0.01 	&	0.029 	&	9.52 	&	9.59 	&	17.37 	&	3.12 	\\
363569997	&	F0	&	7393 	$\pm$	16 	&	0.06 	$\pm$	0.01 	&	3.96 	$\pm$	0.02 	&	0.025 	&	17.29 	&	17.44 	&	34.83 	&	5.54 	\\
364824517	&	A6	&	7587 	$\pm$	14 	&	-0.67 	$\pm$	0.01 	&	4.28 	$\pm$	0.02 	&	0.023 	&	19.88 	&	27.40 	&	30.10 	&		\\
364827308	&	A7	&	7700 	$\pm$	9 	&	-0.12 	$\pm$	0.01 	&	4.02 	$\pm$	0.01 	&	0.023 	&	18.26 	&	20.22 	&	26.39 	&	4.21 	\\
365069183	&	A6	&	7206 	$\pm$	12 	&	-0.13 	$\pm$	0.01 	&	4.13 	$\pm$	0.01 	&	0.015 	&	15.56 	&	33.22 	&	30.58 	&	5.06 	\\
365973747	&	A6	&	7758 	$\pm$	9 	&	-0.42 	$\pm$	0.01 	&	4.14 	$\pm$	0.01 	&	0.018 	&	23.08 	&	28.62 	&	27.32 	&	5.54 	\\
366419043	&	F0	&	7577 	$\pm$	7 	&	-0.07 	$\pm$	0.01 	&	3.93 	$\pm$	0.01 	&	0.027 	&	11.92 	&		&		&		\\
369471949	&	F0	&	7024 	$\pm$	38 	&	0.17 	$\pm$	0.02 	&	3.84 	$\pm$	0.05 	&	0.023 	&		&		&		&		\\
369535670	&	A7	&	7673 	$\pm$	9 	&	-0.16 	$\pm$	0.01 	&	4.01 	$\pm$	0.01 	&	0.011 	&	22.31 	&	45.16 	&	39.06 	&	6.37 	\\
383452490	&	A6	&	7743 	$\pm$	18 	&	-0.27 	$\pm$	0.02 	&	4.12 	$\pm$	0.03 	&	0.027 	&	20.83 	&	20.90 	&	40.44 	&	6.75 	\\
386154220	&	A7	&	7732 	$\pm$	14 	&	-0.13 	$\pm$	0.01 	&	4.05 	$\pm$	0.02 	&	0.013 	&	18.08 	&	36.86 	&	35.26 	&	5.71 	\\
386264352	&	A7	&	7436 	$\pm$	7 	&	-0.14 	$\pm$	0.01 	&	4.01 	$\pm$	0.01 	&	0.027 	&	18.03 	&	18.03 	&	25.55 	&	5.75 	\\
386494520	&	A7	&	7405 	$\pm$	85 	&	0.14 	$\pm$	0.08 	&	3.9 	$\pm$	0.1 	&	0.012 	&	18.75 	&	24.90 	&	22.00 	&	3.07 	\\
386523861	&	A6	&	7351 	$\pm$	12 	&	-0.08 	$\pm$	0.01 	&	4.02 	$\pm$	0.02 	&	0.027 	&	14.65 	&	16.50 	&	18.16 	&		\\
387278923	&	A5	&	7816 	$\pm$	14 	&	-0.31 	$\pm$	0.01 	&	4.11 	$\pm$	0.02 	&	0.012 	&	16.43 	&	41.17 	&	38.22 	&	5.89 	\\
388811781	&	F0	&	7008 	$\pm$	130 &	-0.2 	$\pm$	0.1 	&	4.0 	$\pm$	0.2 	&	0.024 	&	10.44 	&	13.56 	&	14.71 	&		\\
389355807	&	A7	&	7961 	$\pm$	21 	&	0.20 	$\pm$	0.02 	&	3.85 	$\pm$	0.03 	&	0.011 	&	23.40 	&	27.98 	&	27.32 	&	4.57 	\\
389356946	&	F0	&	7103 	$\pm$	52 	&	-0.22 	$\pm$	0.04 	&	4.02 	$\pm$	0.07 	&	0.036 	&	9.53 	&	10.23 	&	12.25 	&	2.91 	\\
393054236	&	A7	&	7227 	$\pm$	17 	&	-0.49 	$\pm$	0.01 	&	4.16 	$\pm$	0.02 	&	0.029 	&	17.40 	&	17.64 	&	20.44 	&	5.18 	\\
408237375	&	F0	&	7092 	$\pm$	16 	&	0.22 	$\pm$	0.01 	&	3.81 	$\pm$	0.02 	&	0.025 	&	9.80 	&	9.90 	&	12.89 	&	3.26 	\\
411278907	&	F0	&	7375 	$\pm$	12 	&	-0.05 	$\pm$	0.01 	&	3.93 	$\pm$	0.02 	&	0.023 	&	11.66 	&	25.79 	&	19.00 	&	4.47 	\\
414922359	&	F0	&	7146 	$\pm$	21 	&	0.38 	$\pm$	0.01 	&	3.86 	$\pm$	0.03 	&	0.021 	&	16.89 	&	17.13 	&	34.06 	&	6.24 	\\
417654237	&	A7	&	7346 	$\pm$	8 	&	-0.15 	$\pm$	0.01 	&	3.99 	$\pm$	0.01 	&	0.013 	&	12.29 	&	30.34 	&	24.69 	&	4.69 	\\
417658371	&	F0	&	7211 	$\pm$	9 	&	0.02 	$\pm$	0.01 	&	4.11 	$\pm$	0.01 	&	0.031 	&	12.64 	&	14.52 	&	22.39 	&	4.63 	\\
427738626	&	F0	&	7606 	$\pm$	13 	&	-0.11 	$\pm$	0.01 	&	3.96 	$\pm$	0.02 	&	0.019 	&	17.64 	&	30.69 	&	37.26 	&		\\
429990941	&	A5	&	7912 	$\pm$	9 	&	-0.19 	$\pm$	0.01 	&	4.01 	$\pm$	0.01 	&	0.013 	&	14.59 	&	35.45 	&	28.76 	&	4.43 	\\
434385299	&	F0	&	7558 	$\pm$	24 	&	0.02 	$\pm$	0.02 	&	3.97 	$\pm$	0.03 	&	0.021 	&	19.78 	&	21.67 	&	36.92 	&	6.16 	\\
437045706	&	A7	&	7790 	$\pm$	15 	&	0.00 	$\pm$	0.01 	&	3.93 	$\pm$	0.02 	&	0.012 	&	24.81 	&	31.24 	&	30.37 	&	5.92 	\\
439859762	&	A7	&	7548 	$\pm$	14 	&	-0.06 	$\pm$	0.01 	&	4.01 	$\pm$	0.02 	&	0.022 	&	22.02 	&	22.11 	&	31.83 	&	6.23 	\\
443853630	&	F0	&	7209 	$\pm$	19 	&	0.17 	$\pm$	0.01 	&	3.93 	$\pm$	0.02 	&	0.028 	&	6.61 	&	13.25 	&	17.28 	&	4.23 	\\
444986428	&	A3	&	7792 	$\pm$	10 	&	-0.12 	$\pm$	0.01 	&	3.95 	$\pm$	0.02 	&	0.022 	&	19.14 	&	19.49 	&	28.46 	&		\\
445021749	&	F0	&	7096 	$\pm$	13 	&	-0.16 	$\pm$	0.01 	&	3.99 	$\pm$	0.02 	&	0.035 	&	9.70 	&	10.44 	&	15.64 	&	3.21 	\\
445842085	&	A7	&	7473 	$\pm$	52 	&	-0.29 	$\pm$	0.05 	&	4.16 	$\pm$	0.09 	&	0.027 	&	19.96 	&	20.32 	&	24.71 	&	5.72 	\\
446413092	&	F3	&	6928 	$\pm$	46 	&	0.44 	$\pm$	0.04 	&	3.59 	$\pm$	0.06 	&	0.016 	&	6.72 	&	6.78 	&	10.32 	&	2.20 	\\
453383992	&	A6	&	7279 	$\pm$	28 	&	-0.15 	$\pm$	0.02 	&	4.12 	$\pm$	0.04 	&	0.032 	&	5.38 	&	5.99 	&	21.30 	&		\\
453445162	&	F0	&	6978 	$\pm$	17 	&	0.13 	$\pm$	0.01 	&	3.76 	$\pm$	0.02 	&	0.032 	&	7.89 	&	7.94 	&		&		\\
457037715	&	A6	&	8113 	$\pm$	21 	&	-0.17 	$\pm$	0.02 	&	3.92 	$\pm$	0.03 	&	0.021 	&	5.99 	&	19.58 	&	21.16 	&		\\
457061109	&	F0	&	7298 	$\pm$	25 	&	0.18 	$\pm$	0.02 	&	3.85 	$\pm$	0.03 	&	0.024 	&	6.33 	&	8.72 	&	8.00 	&		\\
458409283	&	F0	&	7042 	$\pm$	39 	&	0.21 	$\pm$	0.03 	&	3.82 	$\pm$	0.06 	&	0.020 	&	9.13 	&	9.18 	&	13.98 	&	3.15 	\\
458459450	&	A7	&	7604 	$\pm$	24 	&	-0.07 	$\pm$	0.02 	&	4.07 	$\pm$	0.03 	&	0.025 	&	18.89 	&	39.70 	&	36.99 	&	6.52 	\\
458650370	&	A6	&	7200 	$\pm$	14 	&	-0.43 	$\pm$	0.01 	&	4.34 	$\pm$	0.02 	&	0.015 	&	41.53 	&	52.23 	&	47.08 	&	7.33 	\\
468985290	&	A7	&	7430 	$\pm$	14 	&	-0.30 	$\pm$	0.01 	&	4.05 	$\pm$	0.02 	&	0.013 	&	15.72 	&	32.44 	&	29.30 	&	5.15 	\\

\hline
\end{longtable}

\appendix
\section{Tables}

\begin{landscape}
\end{landscape}

\end{document}